\documentclass[journal]{IEEEtran}
\usepackage{cite}
\usepackage{amsmath,amssymb,amsfonts}
\usepackage{algorithmic}
\usepackage{graphicx}
\usepackage{textcomp}
\usepackage{harpoon}
\usepackage{amsmath}
\usepackage{mathrsfs}
\usepackage{amssymb}
\usepackage{epstopdf}
\usepackage{subfig}
\usepackage{lmodern}
\usepackage{bm}
\usepackage{sansmath}
\usepackage{booktabs}
\usepackage{colortbl}
\usepackage{esint}
\usepackage{stfloats}
\usepackage{lipsum}
\usepackage{multicol}
\usepackage{multirow}
\usepackage{tabularx}
\usepackage{color}
\usepackage{cases}

\newenvironment{proof}{{\textsl {Proof:}}}{\hfill$\blacksquare$\par}
\ifCLASSINFOpdf

\else

\fi

\usepackage[absolute,showboxes]{textpos}

\TPMargin{3pt}

\newcommand{\copyrightstatement}{
	\begin{textblock}{0.84}(0.08,0.95) 
		\noindent
		\footnotesize
		\copyright 2023 IEEE. Personal use of this material is permitted. Permission from IEEE must be obtained for all other uses, in any current or future media, including reprinting/republishing this material for advertising or promotional purposes, creating new collective works, for resale or redistribution to servers or lists, or reuse of any copyrighted component of this work in other works. DOI: 10.1109/TIFS.2023.3240028.
	\end{textblock}
}

\begin{document}
\copyrightstatement
\title{Robust Multi-Beam Secure mmWave Wireless Communication for Hybrid Wiretapping Systems}

\author{Bin~Qiu,~\IEEEmembership{Member,~IEEE,}
	Wenchi~Cheng,~\IEEEmembership{Senior Member,~IEEE,}
	and~Wei~Zhang,~\IEEEmembership{Fellow,~IEEE}
	\thanks{This work was supported in part by the National Key Research and Development Program of China under Grant 2021YFC3002102, in part by the Key R\&D Plan of Shaanxi Province under Grant 2022ZDLGY05-09, in part by the Fundamental Research Funds for the Central Universities under Grant XJS220105, in part by the Project funded by China Postdoctoral Science Foundation under Grant 2022M712491, and in part by the Natural Science Basic Research Program of Shaanxi under Grant 2023-JC-QN-0715. (Corresponding author: Wenchi Cheng.)
	}
	\thanks{Bin Qiu and Wenchi Cheng are with the State Key Laboratory of Integrated Services Networks, Xidian University,	Xian 710071, China (e-mail: qiubin@xidian.edu.cn; wccheng@xidian.edu.cn).}
	\thanks{Wei Zhang is with the School of Electrical Engineering and Telecommunications, University of New South Wales, Sydney, NSW 2052, Australia (e-mail: w.zhang@unsw.edu.au).}

}

\maketitle

\begin{abstract}
In this paper, we consider the physical layer (PHY) security problem for hybrid wiretapping wireless systems in millimeter wave transmission, where active eavesdroppers (AEs) and passive eavesdroppers (PEs) coexist to intercept the confidential messages and emit jamming signals. To achieve secure and reliable transmission, we propose an artificial noise (AN)-aided robust multi-beam array transceiver scheme. Leveraging beamforming, we aim to minimize transmit power by jointly designing the information and AN beamforming, while satisfying valid reception for legitimate users (LUs), per-antenna power constraints for transmitter, as well as all interception power constraints for eavesdroppers (Eves). In particular, the interception power formulation is taken into account for protecting the information against hybrid Eves with imperfect AE channel state information (CSI) and no PE CSI. In light of the intractability of the problem, we reformulate the considered problem by replacing non-convex constraints with tractable forms. Afterwards, a two-stage algorithm is developed to obtain the optimal solution. Additionally, we design the received beamforming weights by means of minimum variance distortionless response, such that the jamming caused by AEs can be effectively suppressed. Simulation results demonstrate the superiority of our proposed scheme in terms of energy efficiency and security.
\end{abstract}


\IEEEpeerreviewmaketitle

\section{Introduction}

\IEEEPARstart{I}{n} wireless communication, due to the ever increasing demands in data rate, many advanced techniques have attracted more attention \cite{OAM-NFC, OAM}. Specifically, millimeter wave (mmWave) transmission provides multigigabits per second data rate due to its inherent advantages such as very wide bandwidth \cite{Millimeter-wave-Ming}. It has been recognized as a promising technique for next generation wireless networks \cite{Sub-6-GHz-Omid, Ultra_Low}. However, mmWave transmission is subject to a large path loss and ensues high noise levels for the small wavelengths in mmWave band. Most mmWave communication systems employ beamforming transmission based on phased-array technique in providing array gain to enhance the coverage distance \cite{Physical-Layer-Security-Shu}. Along with a number of applications of mmWave communication, its security is tremendous priority but challenging, as the openness nature allows undesired users to wiretap the confidential messages. In addition to upper-layer encryption, physical layer (PHY) security has attracted considerable attention over the past decade \cite{A-survey-Zou}. In PHY security of wireless communication, the main idea is to exploit the intrinsic nature of wireless medium to realize transmission of information from transmitter to intended users, while avoiding the information leakage \cite{Ultra-Reliable}. 

It is of interest to develop beamforming techniques that provide an extra security by exploiting flexibility at PHY. As a pioneer on PHY security, Wyner introduced wiretap channel model and laid a foundation of information theory of PHY security \cite{wire-tap-channel-Wyner}. Thereafter, Csis\'{z}ar and K\"{o}rner generalized the degraded wiretap channel to broadcast channel with confidential messages, and analysed the secrecy capacity of a more general wiretap channel \cite{Broadcast-channels-Csiszar}. Based on these foundation works, there has been growing interest in the secure transmission using beamforming techniques. Inspired by criterion of secrecy, the authors of \cite{Opportunistic-Scheduling-Abbas} aimed to directly maximize the secrecy rate for PHY security enhancement. Focusing on mmWave communication systems, the authors of \cite{Millimeter-Wave-Systems,Randomly-Located-Eavesdroppers} employed maximum ratio transmitting beamforming scheme to guarantee secure transmission. The authors of \cite{Polygon-Zhang} proposed a phased-array transmission structure via polygon construction PSK modulation to achieve secure transmission. The authors of \cite{Hybrid-MIMO-Wang} developed a hybrid phased multiple-input multiple-output (MIMO) secure mmWave wireless communications scheme, where flexible phased-MIMO antenna enjoyed the advantages of MIMO in spatial diversity without sacrificing the main advantage of phased-array in coherent directional gain.

Another efficient PHY way for guaranteeing secure communication is the embedding of artificial noise (AN), also called as jamming noise. Noteworthy, the construct of AN-aided beamforming was first introduced in \cite{Guaranteeing-Secrecy-Goel}. AN, which is generated by consuming a certain transmit power, is transmitted simultaneously with information for the purpose of hiding information transmission \cite{Frequency_Diverse_Array}. Inspired by this, the authors of \cite{Covert-Communication-Chen} presented an AN-assisted interference alignment with wireless power transfer scheme. The transmission rate was maximized by jointly optimizing the information transmit power and the coefficient of power splitting. Additionally, the authors of \cite{MISOME-Wiretap} investigated on-off transmission and adaptive transmission schemes for multiple-input, single-output, and multi-antenna eavesdropper (MISOME) wiretap channels. In practice, perfect knowledge of eavesdropper (Eve) cannot be acquired by the transmitter. To withstand the imperfect knowledge, several robust synthesis schemes of secure transmission were developed, e.g., main-lobe-integration design \cite{Robust-Secure-Transmission-Shu} and outage-constrained design \cite{Proximal-Security-Lin}.

Most of previous works on PHY security focused on passive eavesdropping. While the transmitter and legitimate user (LU) take full advantage of secrecy enhancement techniques to ensure the valid reception of LU and avoid wiretapping, the Eve also adopts advanced technologies to compromise the security of communication \cite{hybrid-adversary-Basciftci,active-eavesdropper-Wu,Full-Duplex-Hybrid-Attacker-Ahuja}. In this aspect, the Eve can actively jam and attack, called as active eavesdroppers (AEs), to deteriorate the reception of LU. In \cite{Secrecy-Performance-Allipuram}, the secrecy performance of wiretap channels in the presence of perfect channel state information (CSI) of AE was examined. In practice, different types of Eves coexist, which cause severe secrecy and reliability performance degradation. Even in some hybrid wiretapping scenarios, Eves can randomly switch between passive wiretapping and active jamming \cite{Random-Matrix-Theory-Chen}.

Motivated by above-mentioned works, in this paper we attempt to achieve robust multi-beam secure transmission in the presence of hybrid wiretapping environment. More specifically, leveraging beamforming, we aim to minimize the total transmit power by jointly designing the transmit information and AN beamforming for the purpose of reducing message leakage, subject to prescribed received signal-to-interference-plus-noise ratio (SINR) requirements, per-antenna transmit power constraints, as well as interception power constraints. Realistically, passive eavesdropper (PE) usually keeps silent to hide their existence. The PE tries to intercept the confidential messages without emitting signals. Therefore, the information of PE is not available by the transmitter. The AE, as a potential Eve who attempts to intercept the confidential messages and jam LUs’ reception, can be discovered by the transmitter due to the emission of jamming signals. Considering some uncontrollable factors of AEs, such as dynamical movement or random switching between passive wiretapping and active jamming, the information of AE is also hard to be precisely got by the transmitter. As a result, our design is under practical cases of uncertain AE CSI and no PE CSI. Different from the typical sum-power constraint, we consider a more realistic per-antenna power constraint. Each antenna is equipped with its own power amplifier. Thus, per-antenna peak power is confined individually within the linearity of the power amplifier in the practical implementation. In what follows, the weight vectors at LUs are designed by the minimum variance distortionless response (MVDR) method, which is capable of controlling the received beamforming for suppressing the jamming from AEs. 

The transmit power minimization problem is challenging due to the non-concave and the probability constraints. To solve the considered problem, we employ a two-stage procedure. In the first stage, we solve the transmit power minimization problem with the fixed SINR tolerance. More specifically, the worst-case received SINR constraint is recast as a finite number of constraints by applying {\textit{S-Lemma}}\cite{Convex-optimization-Boyd}. The probabilistic outage constraint is transformed into a convex performance lower bound. Then, the reformulated problem is in a form suitable for semidefinite programming (SDP). Next, in the second stage, we aim to minimize the eavesdropping rate by adjusting the SINR tolerance. The exponential penalty method \cite{exponential-penalty-Li} is adopted to transform the optimization problem, followed by a quasi-Newton algorithm referred to as the Broyden-Fletcher-Goldfarb-Shanno (BFGS) algorithm \cite{BFGS-Nezhad} to obtain the global optimal solutions.

Hereafter, the paper is organized as follows. Section II gives the system model. Section III derives the optimization problem for the array transmit design. In Section IV, we develop a two-stage algorithm to solve the considered problem. Simulation results are presented in Section V. Finally, we conclude in Section VI.

\textsl{Notations}: Matrices, vectors, and scalars are denoted by bold upper-case, bold lower-case, and lower case letters, respectively. Superscripts $( \cdot )^ {-1} $, $( \cdot )^T$, and $( \cdot )^H$ denote the inverse, transpose, and Hermitian, respectively. The expectation, trace, and rank of matrix are denoted by $\mathbb{E}\{\cdot\}$, Tr$(\cdot)$, and Rank$(\cdot)$, respectively. We use $\buildrel \Delta \over = $ to denote definition operations. We use ${[ \cdot ]_{i}}$ and ${[ \cdot ]_{i,j}}$ to indicate the $i$th element of the vector and the entry in the $i$th row and the $j$th column of a matrix, respectively. We use vec$(\cdot)$ to stack columns of matrix into a single column vector. The $\ell_2$-norm and modulus are denoted by $\left\|  \cdot  \right\|_2$ and $\left|  \cdot  \right|$, respectively. The maximum eigenvalue and the $i$th eigenvalue of a matrix are denoted by ${\lambda _{\max }}\left(  \cdot  \right)$ and ${\lambda _{i}}\left(  \cdot  \right)$, respectively. We use ${\mathfrak{Re}}\{\cdot\}$ and ${\mathfrak{Im}}\{\cdot\}$ to extract the real valued part and imaginary valued part of a complex number, respectively. Matrices ${{\mathbf{I}}_N}$ and ${{\boldsymbol{0}}_{N \times M}}$ stand for the identity matrix with $N \times N$ and zeros matrix with $N \times M$, respectively. We use $\mathbb{R}^{N \times M}$ and $\mathbb{C}^{N \times M}$ to represent $N \times M$ real and complex entries, respectively. We denote by ${\nabla _{\bf{x}}}f\left(  \cdot  \right)$ the gradient of $f\left(  \cdot  \right)$ with respect to ${\bf{x}}$. The $N \times N$ complex Hermitian matrix is denoted by $\mathbb{H}^N$.

\section{System Sketch and Signal Model}

\begin{figure}
	\centering
	\includegraphics[width=0.85\columnwidth]{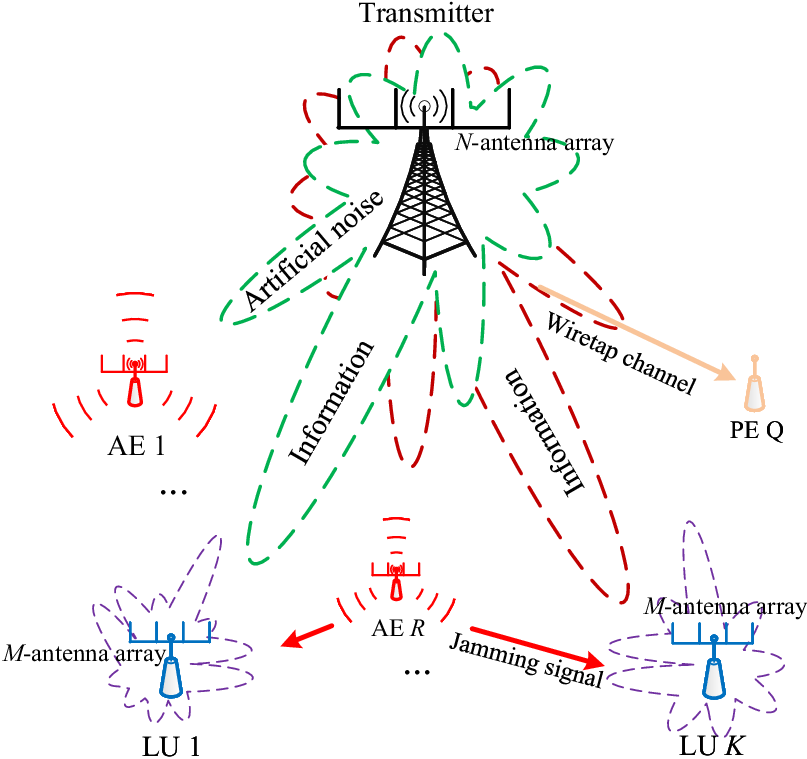}
	\caption{Generic system model of the secure beamforming with hybrid eavesdroppers.}
	\label{fig1}
\end{figure}

Let us consider a line of sight (LoS) mmWave multi-user multiple input multiple output (MU-MIMO) transmission system, as shown in Fig. \ref{fig1}, which consists of a
 transmitter and $K$ LUs in the presence of hybrid Eves. The transmitter is equipped with $N$-antenna array, and each LU is equipped with $M$-antenna array. Unless otherwise stated, antenna array is assumed as a uniform linear array (ULA), and the results can be easily extended to multidimensional periodic arrays. The first antenna of transmit array and  the first antenna of each LU are chosen as reference elements. It is required that the transmitter forwards multiple message streams to corresponding LUs. In the hybrid wiretapping system, we assume $R$ AEs and $Q$ PEs coexist to intercept the confidential messages and/or emit jamming signals to compromise the links between the transmitter and the LUs. For mmWave beamforming transmission, both far-field parallel wavefront and LoS channel can hold \cite{backhaul-access-Maltsev}.

The channel vector$\footnote{{The channel model can be extended to the Saleh-Valenzuela geometric model with multi-path transmission following the spirit of \cite{Polygon-Zhang}.}}$ for transmission and steering vector for reception along the spatial location $(r, \theta)$ in the free-space path loss model, denoted by ${{\bf{h}}}(r,\theta)$ and ${{\bf{a}}}(\theta)$, can be given by \cite{Hybrid-MIMO-Wang,Proximal-Security-Lin} 
\begin{eqnarray}
\begin{aligned}[b]
	{{\bf{h}}}(r, \theta)  =  {\rho}(r) {\left[ {1,{e^{ j2\pi \frac{{{f_c}{d_t}\sin \theta }}{c}}},...,{e^{ j2\pi \frac{{{f_c}\left( {{N} -1} \right){d_t}\sin \theta }}{c}}}} \right]^H},
\end{aligned}
\label{eq1}\end{eqnarray}
and
\begin{eqnarray}
\begin{aligned}[b]
	{{\bf{a}}}(\theta) =  {\left[ {1,{e^{j2\pi \frac{{{f_c}{d_l}\sin \theta }}{c}}},...,{e^{j2\pi \frac{{{f_c}\left( {M -1} \right){d_l}\sin \theta }}{c}}}} \right]^H},
\end{aligned}
\label{eq2}\end{eqnarray}
respectively, where $c$ is the speed of light, $r$ denotes the distance between the transmitter and the receiver, $\theta$ indicates the direction of the receiver relative to the transmitter, $f_c$ denotes the carrier frequency, $d_t$ and $d_l$ refer to the inter-element spacing of the ULA at the transmit array and LUs, respectively, and $\rho \left( {{r}} \right)$ denotes the signal attenuation factor. For the far-field transmission model, i.e., $r \gg d_t$, the attenuation difference among the transmit antennas can be ignored.

\begin{figure}[tb]
	\begin{centering}
		\subfloat[]{\begin{centering}
				\includegraphics[width=0.7\columnwidth]{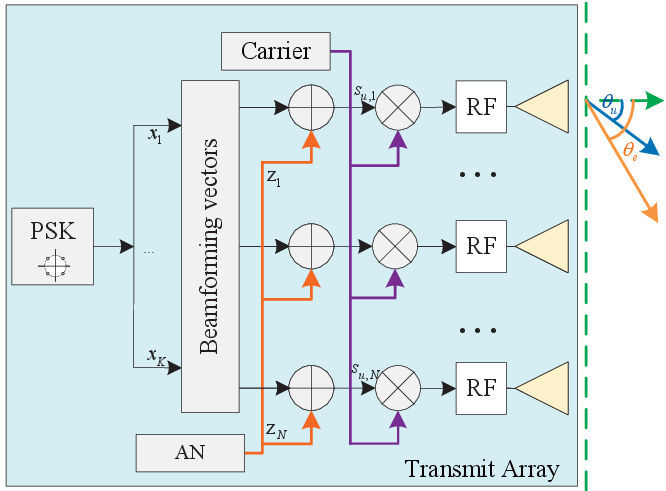}
				\par\end{centering}
		}\\
		\subfloat[]{\begin{centering}
				\includegraphics[width=0.6\columnwidth]{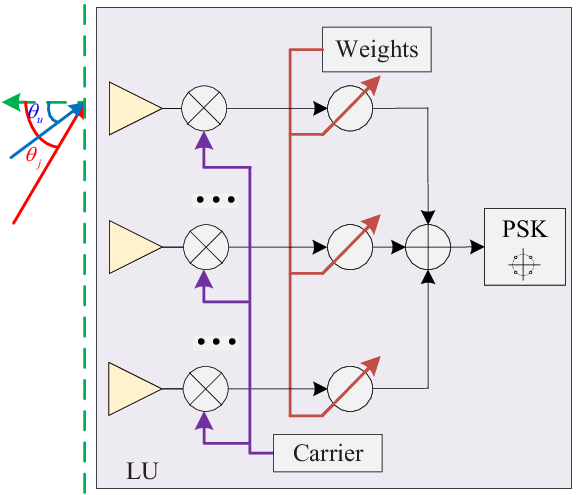}
				\par\end{centering}
		}
		\par\end{centering}
	\caption{Signal architecture for the array transceiver structure. (a) transmission architecture. (b) received architecture.}
	\label{fig2}\end{figure}

Based on the strategy of array transmit structure, as shown in Fig. \ref{fig2}(a), the AN-aided multi-beam transmit baseband signal, denoted by ${{\bf{s}}_u}$, is given by
\begin{eqnarray}
\begin{aligned}[b]
	{{\bf{s}}_u} = \underbrace{\sum\limits_{k \in \mathcal{K}} {\bf{w}}_k{x_k}}_{{\text{Information streams}}} + \underbrace{{\bf{z}}}_{{\text{AN}}},
\end{aligned}
\label{eq3}\end{eqnarray}
where ${{{\bf{w}}_k}}\in \mathbb{C}^{N \times 1}$ is the beamforming vector for message $x_k$, satisfying $\mathbb{E}[{\left| {x_k} \right|^2}] = 1$, which is the modulated symbol to LU $k$, $\forall k \in \mathcal{K}$, $\mathcal{K}  =  \left\{ {1,2,...,K} \right\}$, and ${{\mathbf{z}}}$ is the AN whose elements consist of zero-mean and variance ${\bf{Z}}$ complex Gaussian random variables, i.e., ${\bf{z}}\sim{\mathcal{CN}}({{{0}}},{\bf{Z}})$, ${\bf{Z}} \in \mathbb{H}^{N}$, and ${\bf{Z}}\succeq  \boldsymbol{0}$.  

It is noted that LUs send handshaking/beacon signals to the transmitter to report their status (service requirements) \cite{Wireless-Information-and-Power-Transfer-Ng}. In particular, the corresponding transmitter-to-LU directional information can be acquired with the direction of arrival (DoA) algorithms \cite{DOA-estimation-Shu}. Therefore, we assume that the perfect CSI of LUs can be obtained at the transmitter before initiating secure transmissions. Denote by $r_{u,k}$, and $\theta_{u,k}$ the range and direction of LU $k$ related to the transmitter, $k \in \mathcal{K}$, respectively. We employ the phased array structure at LUs, as shown in Fig. \ref{fig2}(b), to eliminate the jamming emitted by AEs. The jamming signals from AEs are the transmission of uninterrupted high-power noise signals to form a suppression interference environment in the space. In addition, we assume that the jamming signals emitted by AEs are noise signals with the same frequency as the carrier from transmitter, otherwise LUs can eliminate the jamming signals by filtering in frequency domain. To simplify the notations, we use ${{\bf{h}}_{u,k}}$, ${{\bf{a}}_{u,k}}$, and ${{\bf{a}}_{j,k,r}}$ to denote the LU channel vector, steering vector of LU, and steering vector of jamming, i.e., ${{\bf{h}}_{u,k}} ={{\bf{h}}}(r_{u,k}, {\theta _{u,k}})$, ${{\bf{a}}_{u,k}} ={{\bf{a}}}({\theta _{u,k}})$, and ${{\bf{a}}_{j,k,r}} ={{\bf{a}}}({\theta _{j,k,r}})$ with ${\theta _{j,k,r}}$ being the jamming direction of AE $r$ related to LU $k$. Assuming that all transceivers process with ideal time and frequency synchronization, the received down-converted signal at LU $k$, $k \in \mathcal{K}$, denoted by $y_{u,k}$, can be expressed as follows:
\begin{align}
y_{u,k} =\underbrace {{{\bf{h}}_{u,k}^H}{{\bf{s}}_u}{{\bf{a}}_{u,k}^H}{{\bf{v}}_k}}_{{\text{Information signal}}} + \underbrace {{{\bf{a}}_{j,k}^H}{{\bf{v}}_k}}_{{\text{Jamming signal}}} + \underbrace {{n_{u,k}}}_{{\text{Noise}}},
\label{eq4}\end{align}
where ${{\bf{a}}_{j,k}} = \sum\limits_{r \in \mathcal{R}} {{\rho _{k,r}}\sqrt {{E_{j,r}}} {s_{j,r}}} {{\bf{a}}_{j,k,r}}$, $\mathcal{R} =  \left\{ {1,2,...,R} \right\}$, ${s_{j,r}}$ denotes the jamming signal emitted from AE $r$ with $\mathbb{E}[{\left| {{s_{j,r}}} \right|^2}] = 1$, ${E_{j,r}}$ is transmit jamming power of AE $r$, ${\rho_{k,r}}$ is the path loss factor between AE $r$ and LU $k$, ${\bf{v}}_{k}\in \mathbb{C}^{M \times 1}$ is the weight vector of LU $k$, and ${{{n}}_{u,k}}$ is the complex additive white Gaussian noise (AWGN) with zero mean and variance $\sigma _{u,k}^2$, i.e., ${{{{n}}_{u,k}}}\sim{\mathcal{CN}}({{{0}}},\sigma _{u,k}^2)$.

Due to the emission,  AEs can be discovered, and thus part CSI is obtained by the transmitter. Nevertheless, considering the uncontrollable factors of AEs, such as random switching between passive wiretapping and active jamming, as well as dynamical movement, AEs should be viewed as potential Eves and there are always some uncertainty or measured errors in the CSI of AEs. On account of this, the beamforming design is not only required to prevent eavesdropping from AE, but need to achieve robust transmission with imperfect CSI of AEs. We develop a robust secure beamforming scheme via a moment-based random model, where only some roughly estimated first and second order statistics of AEs are available. Therefore, the channel vector from transmitter to AE $r$, $r \in \mathcal{R}$, denoted by ${{\bf{h}}_{a,r}}$, can be expressed as follows:
\begin{align}
{{\bf{h}}_{a,r}} = {{\bf{\hat h}}_{a,r}} + \Delta {{\bf{h}}_{a,r}}, \label{eq5}
\end{align}
where ${{\bf{\hat h}}_{a,r}}$ denotes the CSI estimate of AE $r$ available at the transmitter, and $\Delta {{\bf{h}}_{a,r}}$ denotes the CSI uncertainty of AE $r$. We further assume that the CSI uncertainty with different AEs are independent with each other and have equal variance. We denote by ${\Omega _r}$ a set of all possible CSI uncertainty of AE $r$, which is given by 
\begin{align}
	{\Omega _r}  = \left\{ {\Delta {{\bf{h}}_{a,r}}{\in \mathbb{C}^{N \times 1}}: \Delta {\bf{h}}_{a,r}^H\Delta {{\bf{h}}_{a,r}} \le \varepsilon _r^2} \right\},r \in \mathcal{R},
	\label{eq6}\end{align}
where the radius  $\varepsilon_r>0$ denotes the size of the uncertainty region. We assume that the jamming among AEs can be eliminated through cooperation \cite{cellular-networks-Geraci}. Therefore, the received signal at AE $r$, $r \in \mathcal{R}$, denoted by ${y_{a,r}}$, is given by
\begin{eqnarray}
\begin{aligned}[b]
{y_{a,r}} = {\bf{h}}_{a,r}^H{{\bf{s}}_u} + {n_{a,r}},
\end{aligned}
\label{eq7}\end{eqnarray}
where ${{{n}}_{a,r}}$ denotes the complex AWGN at AE $r$ with zero mean and
variance $\sigma _{a,r}^2$, i.e.,  ${n_{a,r}}\sim{\mathcal{CN}}({{{0}}},\sigma _{a,r}^2)$.

On the other hand, PEs keep silent to hide the existence in general. Consequently, the transmitter cannot acquire any CSI of PE. We assume that PEs are single antenna devices. Our design scenario is under a statistical sense of PE CSI. The received signal at PE $q$, $q \in \mathcal{Q}$, denoted by ${y_{p,q}}$, is given by$\footnote{In practice, PEs usually keep radio silence to avoid exposure. Therefore, it is not feasible to get any information on PEs by the transmitter, including the number of PEs. Nevertheless, by supposing a specific value for $Q$, we make sure that the transmitter can handle up to $Q$ PEs.}$
\begin{eqnarray}
\begin{aligned}[b]
{y_{p,q}} = {\bf{h}}_{p,q}^H{{\bf{s}}_u} + {n_{p,q}},
\end{aligned}
\label{eq8}\end{eqnarray}
where $\mathcal{Q} =  \left\{ {1,2,...,Q} \right\}$, and ${{{n}}_{p,q}}$ is the complex AWGN with zero mean and
variance $\sigma _{p,q}^2$, i.e., ${{{{n}}_{p,q}}}\sim{\mathcal{CN}}({{{0}}},\sigma _{p,q}^2)$.
\section{Robust Beamforming Design for Hybrid Eves}
In what follows, we first give a review of the secrecy capacity for the considered system as the secrecy performance metric. Next, a robust beamforming design is formulated as an optimization problem for minimizing the total transmit power.

\subsection{System Secrecy Performance Metrics}
According to the received signal at LU in \eqref{eq4}, the average SINR for LU $k$, $k \in {{\cal K}}$, denoted by ${\gamma _{u,k}}$, is computed as follows:
\begin{align}
&{\gamma _{u,k}} =\notag \\& \frac{{| {{\bf{h}}_{u,k}^H{{\bf{w}}_k}} {{\bf{a}}_{u,k}^H{{\bf{v}}_k}}|^2}}{{\sum\limits_{i \in {{\cal K}}\backslash k} {| {\bf{h}}_{u,k}^H{{{\bf{w}}_i}}{{\bf{a}}_{u,k}^H{{\bf{v}}_k}} |^2}  + | {{\bf{h}}_{u,k}^H{\bf{za}}_{u,k}^H{{\bf{v}}_k}} |^2 + | {{\bf{a}}_{j,k}^H{{\bf{v}}_k}} |^2 + \sigma _{u}^2}},
\label{eq9}\end{align}
where $\sigma _{u}^2$ is the signal processing noise power at LU$\footnote{Without loss of generality, it is fair to pick up the same values of the thermal noise characteristics for all received users due to similar hardware atchitectures, i.e., $\sigma _{u}^2=\sigma _{u,k}^2$, $\forall k$, $\sigma _{a}^2=\sigma _{a,r}^2$, $\forall r$, and $\sigma _{p}^2=\sigma _{p,q}^2$, $\forall q$.}$.

Under the Gaussian channel, the corresponding achievable rate (bit/s/Hz) of LU $k$, $k \in {{\cal K}}$, denoted by ${R_{u,k}}$, in decoding the message can be expressed as follows:
\begin{align}
{R_{u,k}} = {\log _2}(1 + {\gamma _{u,k}}).
\label{eq10}\end{align}
According to the received signal at AE, the upper bound of the received SINR at AE $r$, $r \in {{\cal R}}$, denoted by ${\gamma _{a,r}}$, is given by \cite{Two-Way-Secure-Feng, Two-Way-Wiretap-Channel-El}
\begin{align}
{\gamma _{a,r}} =\frac{{\sum\limits_{k \in {{\cal K}}}| {{\bf{h}}_{a,r}^H {{{\bf{w}}_k}} } |^2}}{{{{\bf{h}}_{a,r}^H{\bf{Z}}{{\bf{h}}_{a,r}}} + \sigma _{a}^2}}.
\label{eq11}\end{align}
Thus, the achievable rate at AE $r$, $r \in {{\cal R}}$, denoted by ${R_{a,r}}$, can be expressed as follows:
\begin{align}
{R_{a,r}} = {\log _2}(1 + {\gamma _{a,r}}).
\label{eq12}\end{align}
The transmit beamforming is designed under an assumption of unknown PE CSI. As a result, we give the upper bound of the received SINR at PE $q$, $q \in {{\cal Q}}$, denoted by ${\gamma _{p,q}}$, as follows:
\begin{align}
{\gamma _{p,q}} =\frac{{\sum\limits_{k \in {{\cal K}}}| {{\bf{h}}_{p,q}^H {{{\bf{w}}_k}} } |^2}}{{{\bf{h}}_{p,q}^H{\bf{Z}}{{\bf{h}}_{p,q}} + \sigma _{p}^2}}.
\label{eq13}\end{align}
Similarly,  the achievable rate at AE $q$, $q \in {{\cal Q}}$, denoted by ${R_{p,q}}$, can be expressed as follows:
\begin{align}
{R_{p,q}} = {\log _2}(1 + {\gamma _{p,q}}).
\label{eq14}\end{align}
Therefore, the secrecy rate, denoted by ${R_s}$, is defined as \cite{Guaranteeing-Secrecy-Goel}
\begin{align}
{R_s} = {\left[ {\mathop {\min }\limits_{k \in {\cal K}} {R_{u,k}} - \max \left\{ {\mathop {\max }\limits_{r \in {\cal R}} {R_{a,r}},\mathop {\max }\limits_{q \in {\cal Q}} {R_{p,q}}} \right\}} \right]^ + }.
\label{eq15}\end{align}
The secrecy rate ${R_s}$ quantifies the maximum achievable data rate for confidential messages reliable transmission from transmitter to LUs such that Eves are unable to intercept confidential messages even if Eves have infinite computational power for intercepting the signals. 
\subsection{Transmit Beamforming Design}
To achieve PHY security, it is important to guarantee the confidential message is able to be reliably demodulated by LUs, but equally important to avoid interception. To this end, we aim to minimize total transmit power by jointly designing the transmit information and AN beamforming. By this way, less transmit message power means a lower message power leakage to Eves, which can reduce the probability of information being intercepted. Accordingly, the transmit power minimization problem is formulated as follows:
\begin{align}
\textbf{P1:}  \kern 2pt &\mathop {\min }\limits_{{\left\{ {{\bf{w}}_k} \right\}}_{k \in {{\cal K}}}, {\bf{Z}}}  \kern 2pt \sum\limits_{k \in {{\cal K}}}\| { {{\bf{w}}_k} } \|_2^2 + {\rm{Tr}}\left( {\bf{Z}} \right) \tag{16a} \label{eq16a}\\
&{\rm{s.t.}} \kern 2pt {\bf{h}}_{u,k}^H{\bf{Z}}{\bf{h}}_{u,k} = {0}, \forall k \in \mathcal{K},\tag{16b} \label{eq16b}\\
&\kern 15pt	\frac{{|{\bf{h}}_{u,k}^H{{\bf{w}}_k}|^2}}{{\sum\limits_{i \in {{\cal K}} \backslash k}|{\bf{h}}_{u,k}^H {{{\bf{w}}_i}}|^2 + \sigma _{u}^2}} \ge {\zeta _{k}},  \forall k \in \mathcal{K},\tag{16c} \label{eq16c}\\
&\kern 15pt \mathop {\max }\limits_{\Delta {{\bf{h}}_{a,r}} \in {\Omega _r}} \frac{{\sum\limits_{k \in {{\cal K}}}| {{\bf{h}}_{a,r}^H {{{\bf{w}}_k}} } |^2}}{{{{\bf{h}}_{a,r}^H{\bf{Z}}{{\bf{h}}_{a,r}}} + \sigma _{a}^2}} \le {\Gamma _{a,r}}, \forall r \in \mathcal{R},\tag{16d} \label{eq16d}\\
&\kern 15pt \Pr \left( {\mathop {\max }\limits_{q \in{\cal Q}}\kern 2pt  {\frac{{\sum\limits_{k \in {{\cal K}}}| {{\bf{h}}_{p,q}^H {{{\bf{w}}_k}} } |^2}}{{{\bf{h}}_{p,q}^H{\bf{Z}}{{\bf{h}}_{p,q}} + \sigma _{p}^2}} \le {\Gamma _{p}}} } \right) \ge \kappa , \tag{16e} \label{eq16e}\\
&\kern 16pt \sum\limits_{k \in {{\cal K}}} {{{\left[ {{{\bf{w}}_k}{\bf{w}}_k^H} \right]}_{n,n}} } +{\left[ {\bf{Z}} \right]_{n,n}} \le {P_n}, \forall n \in \mathcal{N},\tag{16f} \label{eq16f}\\
&\kern 17pt {\bf{Z}} \succeq  \boldsymbol{0},\tag{16g} \label{eq16g}
\end{align}
where $P_n$ is the power budget for the transmit antenna $n$, $\mathcal{N}=  \left\{ {1,2,...,N} \right\}$, ${\zeta _{k}}$ denotes the prescribed minimum received signal-to-noise ratio (SNR) at each of LU, ${\Gamma _{a,r}}$ and $ {\Gamma _{p}}$ represent the maximum received SINR tolerance for successfully decoding at AE and PE, respectively, and $\kappa$ denotes a probability parameter for providing secure communication. The constraint in \eqref{eq16b} is imposed to project the AN to null space of all LU channels. Conversely, AN is viewed as virtual messages by Eves, and thus it protects the confidential message transmission. The constraint in \eqref{eq16c} is to protect the received signal power for message decoding at LU $k$, so that the received SNR at LU $k$ is above SNR threshold ${{\zeta}_k}$. This constraint guarantees that the achievable rate of LU $k$ is $R_{u,k} \ge {\rm{log}}_2(1+{{\zeta _k}})$. The constraint in \eqref{eq16d} represents that the maximum received SINR at AE $r$ is less than the maximum tolerable received SINR ${\Gamma _{a,r}}$ for a given CSI uncertainty set ${\Omega _r}$, $\forall r \in \mathcal{R}$. To guarantee secure communication, it is set as ${\mathop {\min }\limits_{k \in {\cal K}} \left\{ {\zeta _{k}} \right\} \gg \mathop {\max }\limits_{r \in {\cal R}} \left\{ {{\Gamma _{a,r}}} \right\} > 0}$. When the established optimization problem is feasible and PE does not exist in the system, the formulated problem ensures that the secrecy capacity is bounded below by ${R_s} \ge {\rm{log}}_2(1+\mathop {\min }\limits_{k \in {\cal K}}\{{\zeta _k}\}) -  {\rm{log}}_2(1+ \mathop {\max }\limits_{r \in {\cal R}}\{{\Gamma _{a,r}}\})\ge 0$. The constraint in \eqref{eq16e} picks out the minimum outage requirement for PEs. More specifically, the maximum received SINR among all PEs is required to be smaller than the maximum tolerable received SINR ${\Gamma _{p}}$ with at least probability $\kappa$. Assuming that AE does not exist, constraints \eqref{eq16c} and \eqref{eq16e} together ensure that the secrecy rate is bounded below by ${R_s} \ge {\rm{log}}_2(1+\mathop {\min }\limits_{k \in {\cal K}}\{{\zeta _k}\}) -  {\rm{log}}_2(1+ \mathop {\max }\limits_{q \in {\cal Q}}\{{\gamma _{p,q}}\}) \ge {\rm{log}}_2(1+\mathop {\min }\limits_{k \in {\cal K}}\{{\zeta _k}\}) -  {\rm{log}}_2(1+{\Gamma _{p}})$ with probability $\kappa$. The constraint in \eqref{eq16f} denotes the transmit power constraint for per-antenna at the transmitter. It is important to note that each antenna has its own power amplifier in the analog front-end of a physical implementation array transmitter, and it is necessary to limit each of antenna peak power to process within the linear region of the power amplifier \cite{Per-Antenna-Power-Yu}. Therefore, a power constraint imposed on per-antenna basis is more realistic$\footnote{{In fact, our proposed scheme is also suitable for the sum power constraint by modifying the constraint in \eqref{eq16f} with $\sum\limits_{k \in \mathcal{K}} {{\rm{Tr}} ( {{{\bf{W}}_k}} )}  + {\rm{Tr}} ( {\bf{Z}} ) \le P_{\rm{Tol}}$, $P_{\rm{Tol}}=\sum\limits_{n \in \mathcal{N}}{P_n}$.}}$.  The constraint in \eqref{eq16g} confines matrix ${\bf{Z}}$ to be positive semi-definite Hermitian matrix so as to obtain a valid covariance matrix.

\textsl{Remark 1:} Note that the prescribed minimum received SNR for LU  ${\zeta _{k}}$ in \eqref{eq16c} is determined by quality of service (QoS) assurance for each LU. Whereas the maximum tolerable received SINRs for Eves, ${\Gamma _{a,r}}$ and ${\Gamma _{p}}$, are preset values. We need to find the min-max received SINR tolerance for Eves to maximize the system secrecy rate. In the light of this, we develop a two-stage algorithm for the optimization problem in the next section.
\section{A Two-Stage Algorithm for Optimization Problem}
It can be observed that problem \textbf{P1} is a non-convex optimization problem. Constraint \eqref{eq16d} involves infinite inequality constraints for the continuity of uncertainty sets. Besides, the probabilistic constraint \eqref{eq16e} tightly couples optimization variables and is a non-convex constraint. Therefore, there is no standard method to obtain the optimal solution of the considered optimization problem. To solve the non-convex optimization problem, we convert the infinite number of constraints into an equivalent finite number of constraints and transform the probabilistic constrained problem into a tractable convex constraint. Then, we develop a two-stage algorithm to find the global optimal solutions. More specifically, in the first stage, we aim to optimize over the transmit information and AN beamforming with fixed SINR tolerance; next, in the second stage,  we try to find the min-max received SINR tolerance.
\subsection{Reformulation of the Original Problem}
We first focus on optimal algorithm to solve the problem with fixed received SINR tolerance. Let us stack all of LU channels as follows:
\begin{eqnarray}
\setcounter{equation}{17}
\begin{aligned}[b]
{{\bf{H}}_{u, \rm{Tot}}}{{ \buildrel \Delta \over = }}\left[{\bf{h}}_{u,1},{\bf{h}}_{u,2},...,{\bf{h}}_{u,{K}}\right].
\end{aligned}
\label{eq17}\end{eqnarray}
The AN is designed to place the null space of all LU channels to eliminate the AN interference with LUs, such that 
\begin{eqnarray}
\begin{aligned}[b]
{{\bf{H}}^H_{u, \rm{Tot}}}{{\bf{\Pi }}}={\boldsymbol{0}},
\end{aligned}
\label{eq18}\end{eqnarray}
where ${\bf{Z}} = {\bf{\Pi }}{{\bf{\Pi }}^H}$ with  ${\bf{\Pi}} \in {\mathbb{C}^{ N  \times ( {N - K} )}}$. It is requested to more number of transmit antennas than that of LUs, i.e., $N>K$. The LU channel matrix ${{\bf{H}}_{u, \rm{Tot}}^H}$ is operated with the method of singular-value decomposition (SVD), i.e., 
\begin{eqnarray}
\begin{aligned}[b]
{{\bf{H}}_{u, \rm{Tot}}^H} = {\bf{U}}_u[ {\begin{array}{*{20}{c}}
		\boldsymbol{\Sigma}_u &{\boldsymbol{0}}\end{array}} ]{[ {\begin{array}{*{20}{c}}{{{\bf{V}}_u^{(1)}}}&{{{\bf{V}}_u^{(0)}}}\end{array}} ]^H},
\end{aligned}
\label{eq19}\end{eqnarray}
where ${{{\bf{V}}_u^{(0)}}} \in {\mathbb{C}^{ N  \times \left( {N - K} \right)}}$ consists of the last $(N - K)$ right singular vectors corresponding to zero singular values. Then, let us define ${\bf{\Pi }} \buildrel \Delta \over = {{\bf{\bar V}}}{\bf{\bar \Pi}}$, ${\bf{\bar \Pi}} \in {\mathbb{C}^{\left( {N - K} \right) \times \left( {N - K} \right)}}$, ${{\bf{\bar V}}}={{{\bf{V}}_u^{(0)}}}$. Problem \textbf{P1} can be replaced with respect to $({{{\left\{ {{{\bf{w}}_k}} \right\}}_{k \in {{\cal K}}}},{{\bf{\bar \Pi }}}})$ by
\begin{align}
\textbf{P2:}  \kern 2pt &\mathop {\min }\limits_{{{\left\{ {{{\bf{w}}_k}} \right\}}_{k \in {{\cal K}}}},{{\bf{\bar \Pi }}}}  \kern 2pt\sum\limits_{k \in {{\cal K}}} \| {{\bf{w}}_k } \|_2^2 + {\rm{Tr}}\left( {{\bf{\bar \Pi }}{{{\bf{\bar \Pi }}}^H}} \right) \tag{20a} \label{eq20a}\\
&{\rm{s.t.}} \kern 2pt \frac{{|{\bf{h}}_{u,k}^H{{\bf{w}}_k}|^2}}{{\sum\limits_{i \in {{\cal K}} \backslash k} |{\bf{h}}_{u,k}^H{{{\bf{w}}_i}} {|^2} + \sigma _{u}^2}} \ge {\zeta _k},  \forall k \in \mathcal{K}, \tag{20b} \label{eq20b}\\
&\kern 14pt \mathop {\max }\limits_{\Delta {{\bf{h}}_{a,r}} \in {\Omega _r}} \frac{{\sum\limits_{k \in {{\cal K}}}| {{\bf{h}}_{a,r}^H {{{\bf{w}}_k}} } |^2}}{{{{\bf{h}}_{a,r}^H{{\bf{\bar V\bar \Pi }}{{{\bf{\bar \Pi }}}^H}{{{\bf{\bar V}}}^H}}{{\bf{h}}_{a,r}}} + \sigma _{a}^2}} \!\le\! {\Gamma _{a,r}}, \forall r \in \mathcal{R},\tag{20c} \label{eq20c}\\
&\kern 14pt \Pr \left( {\mathop {\max }\limits_{q \in \cal Q}\kern 2pt  {\frac{\sum\limits_{k \in {{\cal K}}}| {{\bf{h}}_{p,q}^H {{{\bf{w}}_k}} } |^2}{{{\bf{h}}_{p,q}^H{{\bf{\bar V\bar \Pi }}{{{\bf{\bar \Pi }}}^H}{{{\bf{\bar V}}}^H}}{{\bf{h}}_{p,q}} + \sigma _{p}^2}} \le {\Gamma _{p}}} } \right) \ge \kappa , \tag{20d} \label{eq20d}\\
&\kern 14pt \sum\limits_{k \in {{\cal K}}} {{{\left[ {{{\bf{w}}_k}{\bf{w}}_k^H} \right]}_{n,n}} } +{\left[ {{\bf{\bar V\bar \Pi }}{{{\bf{\bar \Pi }}}^H}{{{\bf{\bar V}}}^H}} \right]_{n,n}} \le {P_n},\forall n \in \mathcal{N},\tag{20e} \label{eq20e}
\end{align}
Let us define ${{\bf{W}}_k} \buildrel \Delta \over = {{\bf{w}}_k}{\bf{w}}_k^H$ and ${\bf{T}} \buildrel \Delta \over = {\bf{\bar \Pi }}{{\bf{\bar \Pi }}^H}$ to facilitate SDP relaxation.  Problem \textbf{P2} can be converted as an amenable problem, i.e.,
\begin{align}
\textbf{P3:}  \kern 2pt &\mathop {\min }\limits_{{\left\{ {{{\bf{W}}_k}} \right\}}_{k \in {{\cal K}}},{{\bf{T}}}}  \kern 2pt\sum\limits_{k \in {{\cal K}}} {\rm{Tr}} \left({{\bf{W}}_k }\right)  + {\rm{Tr}}\left( {\bf{T}} \right) \tag{21a} \label{eq21a}\\
&{\rm{s.t.}} \kern 2pt 	\frac{{\rm{Tr}}\left({{\bf{H}}_{u,k}{{\bf{W}}_k}}\right)}{{\sum\limits_{i \in {{\cal K}} \backslash k}{\rm{Tr}}\left({\bf{H}}_{u,k}{{{\bf{W}}_i}}\right) + \sigma _{u}^2}} \ge {\zeta _k},  \forall k \in \mathcal{K}, \tag{21b} \label{eq21b}\\
&\kern 14pt \mathop {\max }\limits_{\Delta {{\bf{h}}_{a,r}} \in {\Omega _r}} \frac{{\sum\limits_{k \in {{\cal K}}}{\rm{Tr}}\left( {{\bf{H}}_{a,r}{{{\bf{W}}_k}} }\right)}}{{\rm{Tr}}{\left({\bf{H}}_{a,r}{{\bf{\bar V}}{{\bf{T}}}{{{\bf{\bar V}}}^H}}\right) + \sigma _{a}^2}} \!\le\! {\Gamma _{a,r}}, \forall r \in \mathcal{R},\tag{21c} \label{eq21c}\\
&\kern 14pt \Pr \left( {\mathop {\max }\limits_{q \in \cal Q}\kern 2pt  {\frac{\sum\limits_{k \in {{\cal K}}}{\rm{Tr}}\left( {{\bf{H}}_{p,q}{{{\bf{W}}_k}} }\right)}{{\rm{Tr}}\left({\bf{H}}_{p,q}{{\bf{\bar V}}{{\bf{T}}}{{{\bf{\bar V}}}^H}}\right) + \sigma _{p}^2} \le {\Gamma _{p}}} } \right) \ge \kappa, \tag{21d} \label{eq21d}\\
&\kern 14pt \sum\limits_{k \in \mathcal{K}} {{\rm{Tr}}\left( {{{\bf{E}}^{(n)}}{{\bf{W}}_k}} \right)}  + {\rm{Tr}}\left( {{{{\bf{\bar E}}}^{(n)}}{\bf{T}}} \right) \le {P_n},\forall n \in \mathcal{N},\tag{21e} \label{eq21e}\\
&\kern 16pt {\bf{T}} \succeq  \boldsymbol{0}, \tag{21f} \label{eq21f}\\
&\kern 16pt {{\bf{W}}_k} \succeq  \boldsymbol{0},{\rm{Rank}}\left({{\bf{W}}_k}\right) =1,\forall k \in \mathcal{K}, \tag{21g} \label{eq21g}
\end{align}
where ${{\bf{E}}^{(n)}}  \buildrel \Delta \over = {\bf{i}}_n{\bf{i}}_n^H$, ${{{\bf{\bar E}}}^{(n)}} \buildrel \Delta \over = {\bf{\bar i}}_n{\bf{\bar i}}_n^H$, ${\bf{i}}_n \in {\mathbb{R}^{ N  \times 1}}$ is the $n$th unit vector of length $N$, i.e., ${\bf{i}}_n=[\boldsymbol{0}_{(n-1) \times 1}^T,1,\boldsymbol{0}_{(N-n)\times 1}^T]^T$, ${\bf{\bar i}}_n={\bf{\bar V}}^T[\boldsymbol{0}_{(n-1)\times 1}^T,1,\boldsymbol{0}_{(N-n)\times 1}^T]^T$, ${{\bf{H}}_{a,r}} \buildrel \Delta \over = {{\bf{h}}_{a,r}}{\bf{h}}_{a,r}^H$, ${{\bf{H}}_{p,r}} \buildrel \Delta \over = {{\bf{h}}_{p,q}}{\bf{h}}_{p,q}^H$, ${{\bf{W}}_k} \succeq  \boldsymbol{0}$, ${{\bf{W}}_k} \in {\mathbb{H}^{N}}$, and ${\rm{Rank}}\left({{\bf{W}}_k}\right) =1$ is imposed to guarantee that ${{\bf{W}}_k} \buildrel \Delta \over = {{\bf{w}}_k}{\bf{w}}_k^H$ holds after optimizing. It can be found that constraints in \eqref{eq21b} and \eqref{eq21e} are convex. Constraint \eqref{eq21c} is convex with respect to the optimization variables, but it is a semi-infinite constraint which is intractable. In fact, the following lemma provides a more straightforward method to transform constraint \eqref{eq21c} into a convex linear matrix inequality (LMI) constraint.

\textsl{Lemma 1 (S-Procedure \cite{Convex-optimization-Boyd}):} Let us define a function ${\mathscr{F}}_i\left( {\bf{x}} \right)$, $i \in \{ 1,2\}$ as follows:
\begin{align}
\setcounter{equation}{21}
{\mathscr{F}}_i\left( {\bf{x}} \right)\buildrel \Delta \over = {{\bf{x}}^H}{{\bf{A}}_i}{\bf{x}} + 2{\mathop{{\mathfrak{Re}}}\nolimits} \left\{ {{\bf{b}}_i^H{\bf{x}}} \right\} + {c_i},
\label{eq22}\end{align}
where ${{\bf{A}}_i} \in \mathbb{H}^{N}$, ${\bf{b}}_i \in \mathbb{C}^{N \times 1}$, and ${c_i} \in \mathbb{R}^{1 \times 1}$. Then, the implication ${\mathscr{F}}_1\left( {\bf{x}} \right) \le 0 \Rightarrow {\mathscr{F}}_2\left( {\bf{x}} \right)\le 0$ holds if and only if there exists a variable $\mu>0$, which satisfies
\begin{eqnarray}
\begin{aligned}[b]
\mu \left[ {\begin{array}{*{20}{c}}
	{{{\bf{A}}_1}}&{{{\bf{b}}_1}}\\
	{{\bf{b}}_1^H}&{{c_1}}
\end{array}} \right] \succeq \left[ {\begin{array}{*{20}{c}}
	{{{\bf{A}}_2}}&{{{\bf{b}}_2}}\\
	{{\bf{b}}_2^H}&{{c_2}}
\end{array}} \right].
\end{aligned}
\label{eq23}\end{eqnarray}
Therefore, \eqref{eq23} provides that there exists a point ${{\bf{\hat x}}}$ such that ${\mathscr{F}}_i\left( {\bf{\hat x}} \right)<0$. 

Based on the CSI variance uncertain set of AE in \eqref{eq6}, we have
\begin{align}
{\mathscr{F}}_1\left(\Delta {{\bf{h}}_{a,r}}\right)=\Delta {{\bf{h}}_{a,r}}\Delta {{\bf{h}}_{a,r}} - \varepsilon _r^2 \le 0, r \in \mathcal{R}.
\label{eq24}\end{align}
Then, we insert ${{\bf{h}}_{a,r}} = {{\bf{\hat h}}_{a,r}} + \Delta {{\bf{h}}_{a,r}}$ into constraint \eqref{eq21c} and utilize Lemma 1, yielding
\begin{align}
&{\mathscr{F}}_2\left(\Delta {{\bf{h}}_{a,r}}\right)=\Delta {\bf{h}}_{a,r}^H\left( {\sum\limits_{k \in \mathcal{K}} {\frac{{{{\bf{W}}_k}}}{{{\Gamma _{a,r}}}}}  - {\bf{\bar V T}}{{\bf{\bar V}}^H}} \right)\Delta {{{\bf{h}}}_{a,r}} \notag\\&
+ 2{\mathop{{\mathfrak{Re}}}\nolimits} \left\{ {{\bf{\hat h}}_{a,r}^H\left( {\sum\limits_{k \in \mathcal{K}} {\frac{{{{\bf{W}}_k}}}{{{\Gamma _{a,r}}}}}  - {\bf{\bar V T}}{{\bf{\bar V}}^H}} \right)\Delta {{\bf{h}}_{a,r}}} \right\} \notag\\&
+ {\bf{\hat h}}_{a,r}^H\left( {\sum\limits_{k \in \mathcal{K}} {\frac{{{{\bf{W}}_k}}}{{{\Gamma _{a,r}}}}}  - {\bf{\bar V T}}{{\bf{\bar V}}^H}} \right){{\bf{\hat h}}_{a,r}} - \sigma _{a}^2, r \in \mathcal{R}.
\label{eq25}\end{align}
To conduct ${\mathscr{F}}_2\left(\Delta {{\bf{h}}_{a,r}}\right) \le 0$, there exist ${\mu _r} \ge 0$, ${r \in \mathcal{R}}$, so as to make such following LMI constraint holds, i.e.,
\begin{align}
\begin{array}{l}
{{\bf{S}}_r}\left( {{{\bf{W}}_k},{\bf{T}},{\mu _r}} \right)\\
\kern 4pt	= \left[ {\begin{array}{*{20}{c}}
{{\mu _r}{{\bf{I}}_N} + {\bf{\bar V T}}{{\bf{\bar V}}^H}}&{{\bf{\bar V T}}{{\bf{\bar V}}^H}{{{\bf{\hat h}}}_{a,r}}}\\
{{\bf{\hat h}}_{a,r}^H{\bf{\bar V T}}{{\bf{\bar V}}^H}}&{ - {\mu _r}\varepsilon _r^2 + \sigma _{a}^2 + {\bf{\hat h}}_{a,r}^H{\bf{\bar V T}}{{\bf{\bar V}}^H}{{{\bf{\hat h}}}_{a,r}}}
\end{array}} \right]\\
\kern 4pt	- \frac{1}{{{\Gamma _{a,r}}}}{\bf{U}}_{a,r}^H\sum\limits_{k \in \mathcal{K}} {{{\bf{W}}_k}} {{\bf{U}}_{a,r}} \succeq \boldsymbol{0}, \forall r \in  \mathcal{R},
\end{array}
\label{eq26}\end{align}
where ${{\bf{U}}_{a,r}} = [ {{{\bf{I}}_N},{{{\bf{\hat h}}}_{a,r}}} ]$. One can easily verify that now constraint \eqref{eq21c} has transformed a finite number of constraints such that it facilitates the optimal solutions.

Intuitively, next obstacle to solve problem \textbf{P3} is the probabilistic constraint \eqref{eq21d} with the non-convex tightly coupled optimization variables. To make problem \textbf{P3} a tractable problem, we pursue the constraint \eqref{eq21d} into a convex constraint according to the following lemma. 

\textsl{Lemma 2:} Supposing that the channels of PEs are modeled as independent and identical distributed (i.i.d.) Rayleigh random variables$\footnote{PEs are assumed that can not cooperate and are physically separated by a distance of at least half a wavelength. Furthermore, we suppose that there exist a large number of statistically independent reflected and scattered paths between transmitter and PEs such that these channels construct Rayleigh fading channels \cite{Spatially-Selective-Li} based on the central limit theorem \cite{Fundamentals-WC-Tse}.}$, the probabilistic/chance  constraint \eqref{eq21d} is conservatively transformed as the following constraint:
\begin{align}
&\sum\limits_{k \in \mathcal{K}} {{{\bf{W}}_k}}  - {\Gamma _p}{\bf{\bar VT}}{{{\bf{\bar V}}}^H} \preceq {{\bf{I}}_N}\xi\label{eq27}\\&
\Rightarrow \Pr \left( {\mathop {\max }\limits_{q \in \mathcal{Q}} {\gamma _{p,q}} \le {\Gamma _p}} \right) \ge \kappa,
\label{eq28}\end{align}
where $\xi  = \Phi _N^{ - 1}\left( {1 - {\kappa ^{1/Q}}} \right){\Gamma _p}\sigma _p^2$, with $\Phi _N^{ - 1}\left(  \cdot  \right)$ denoting the inverse cumulative distribution function (c.d.f.) of an inverse central chi-square random variable with $2N$ degrees of freedom (DoF). 

\begin{proof}
Please refer to Appendix A.
\end{proof}

\textsl{Remark 2:} We note that the transformation in \eqref{eq27} can be suitable for any continuous channel distribution by applying an inverse c.d.f. of the corresponding distribution instead of $\Phi _N^{ - 1}\left(  \cdot  \right)$.

\textsl{Remark 3:} It is worth mentioning that the constraint transformation \eqref{eq28} holds but not vice versa since the inequality transforms in \eqref{eq42}. Therefore, \eqref{eq27} is a tightness of \eqref{eq28}.  That is,  replacing constraint \eqref{eq28} with \eqref{eq27} generates a small feasible solution set.

One can easily verify that constraint \eqref{eq27} is tractable in a sense that: i) a feasible solution satisfying \eqref{eq27} is also feasible for constraint \eqref{eq21d}; and ii) the converted constraint in \eqref{eq27} is a convex LMI constraint \cite{Convex-optimization-Boyd}. As a result, replacing constraint \eqref{eq21c} with \eqref{eq26} and substituting \eqref{eq27} for \eqref{eq21d}, problem \textbf{P3} can be recast as follows:
\begin{align}
\textbf{P4:}  \kern 2pt &\mathop {\min }\limits_{{{\left\{ {{{\bf{W}}_k}} \right\}}_{k \in {{\cal K}}}},{{\bf{T}}},{\boldsymbol{\mu}}}  \kern 2pt\sum\limits_{k \in {{\cal K}}} {\rm{Tr}} \left({{\bf{W}}_k }\right)  + {\rm{Tr}}\left( {\bf{T}} \right) \tag{29a} \label{eq29a}\\
&{\rm{s.t.}} \kern 2pt 	\frac{{\rm{Tr}}\left({{\bf{H}}_{u,k}{{\bf{W}}_k}}\right)}{{\sum\limits_{i \in {{\cal K}} \backslash k}{\rm{Tr}}\left({\bf{H}}_{u,k}{{{\bf{W}}_i}}\right) + \sigma _{u}^2}} \ge {\zeta _k},  \forall k \in \mathcal{K}, \tag{29b} \label{eq29b}\\
&\kern 17pt {{\bf{S}}_r}\left( {{{\bf{W}}_k},{\bf{T}},{\mu _r}} \right)\succeq  \boldsymbol{0}, \forall r \in \mathcal{R},\tag{29c} \label{eq29c}\\
&\kern 16pt \sum\limits_{k \in \mathcal{K}} {{{\bf{W}}_k}}  - {\Gamma _p}{\bf{\bar VT}}{{{\bf{\bar V}}}^H} \preceq {{\bf{I}}_N}\xi, \tag{29d} \label{eq29d}\\
&\kern 15pt \sum\limits_{k \in \mathcal{K}} {{\rm{Tr}}\left( {{{\bf{E}}^{(n)}}{{\bf{W}}_k}} \right)}  + {\rm{Tr}}\left( {{{{\bf{\bar E}}}^{(n)}}{\bf{T}}} \right) \le {P_n},\forall n \in \mathcal{N}, \tag{29e} \label{eq29e}\\
&\kern 17pt {\bf{T}} \succeq  \boldsymbol{0}, \tag{29f} \label{eq29f}\\
&\kern 17pt {{\bf{W}}_k} \succeq  \boldsymbol{0}, {\rm{Rank}}\left({{\bf{W}}_k}\right) =1,\forall k \in \mathcal{K}, \tag{29g} \label{eq29g}\\
&\kern 17pt {\mu _r}  \ge 0,\forall r \in \mathcal{R}, \tag{29h} \label{eq29h}
\end{align}
where ${\boldsymbol{\mu}} \buildrel \Delta \over = [{\mu _1},{\mu _2},...,{\mu _R}]^T \in \mathbb{R}^{R \times 1}$ is auxiliary variable vector, whose elements ${\mu _r}>0$, $\forall r \in \mathcal{R}$, were introduced in \eqref{eq26}. An important observation is that problem \textbf{P4} is in a form suitable for SDP relaxation \cite{Quasi-maximum-Wing}; that is, dropping the rank-one constraint ${\rm{Rank}}\left({{\bf{W}}_k}\right) =1$, $\forall k \in \mathcal{K}$, the optimization problem becomes a so-called rank relaxed problem, which is a convex SDP. In this way, the optimization problem can be efficiently solved by using SDP solvers, such as SeDuMi \cite{SeDuMi-Sturm} and SDPT3 \cite{SDPT3-Toh}. According to the basic principle of optimization theory, if the solution ${{\bf{W}}_k}$, $\forall k \in \mathcal{K}$,  of the relaxed problem satisfies a rank-one matrix, it is the optimal solution of the original problem in \textbf{P4}. Crucially, following lemma proves that the rank relaxation is tight.

\textsl{Lemma 3:} The optimal beamforming matrix ${{\bf{W}}_k}$ of \textbf{P4} admits ${\rm{Rank}}\left({{\bf{W}}_k}\right) =1$, $\forall k \in \mathcal{K}$.

\begin{proof}
Please refer to Appendix B.
\end{proof}

\subsection{Secrecy Rate Maximization}
In the previous discussion, we have obtained the optimal transmit information and AN beamforming with considering the SINR tolerance to be fixed. It is observed that we do not maximize the secrecy rate. Next, let us return to adjust the SINR tolerance of Eves to enhance the secrecy performance. To simplify the formulation, we define ${{\mathcal{T}} _{tol}} \buildrel \Delta \over = \{ {\Gamma _{a,1}},{\Gamma _{a,2}},...,{\Gamma _{a,R}},{\Gamma _p}\} $ to denote the SINR tolerance set for Eves. In particular, we denote the SINR tolerance for Eves by ${{\Gamma }_r} \in {\mathcal{T}} _{tol}$, $r \in {{\mathcal{R}} _p}$, ${{\mathcal{R}} _p} = \left\{1,2,...,R,R + 1\right\}$. The optimal problem with respect to the optimization variables ${\Gamma _r}$, $r \in {{\mathcal{R}} _p}$, is then formulated as follows:
\begin{eqnarray}
\setcounter{equation}{30}
\textbf{P5:} \kern 4pt \mathop {\min }\limits_{\left\{{{\Gamma }_r}\right\}}\kern 2pt \mathop {\max }\limits_{r \in {{\mathcal{R}} _p}}  \kern 2pt\mathscr{P} \left({{\Gamma}_r}\right) \kern 4pt {\rm{s.t.}} \kern 2pt \eqref{eq26}, \eqref{eq27}.
\label{eq30}\end{eqnarray}
where $\mathscr{P}\left({{\Gamma}_r}\right) ={{{\log }_2}\left( {1 +{{\Gamma}_r}} \right)}$, $\forall  r \in {\mathcal{R}_p}$. Problem \textbf{P5} is imposed to maximize the secrecy rate by minimizing the maximum Eve achievable rate for a given LU achievable rate, and thus enhance the system secrecy performance.

To solve the min-max problem \textbf{P5}, we resort to the exponential penalty method\cite{exponential-penalty-Li}, which is one of the important optimization methods both in theoretical and algorithmic developments as an instrument for converting constrained problem into unconstrained one. Toward this end, we seek to optimize over $\left\{{{\Gamma }_r}\right\}$, $\forall  r \in {\mathcal{R}_p}$, with fixed $\left({\left\{ {{{\bf{W}}_k}} \right\}}_{k \in {{\cal K}}},{\bf{T}}, {\boldsymbol{\mu}}\right)$. One way to deal with this problem is to transform it into an equivalent nonlinear programming problem, i.e.,
\begin{align}
\textbf{P6:}\kern 4pt	&\mathop {\min }\limits_{{\omega, \left\{{{\Gamma }_r}\right\}}} \kern 2pt \omega \tag{31a} \label{eq31a}\\
& {\rm{s.t.}} \kern 2pt \mathscr{P} \left({{\Gamma}_r}\right)  \le  \omega, \forall  r \in {{\mathcal{R}} _p}, \tag{31b} \label{eq31b}\\
&  \kern 17pt {\bf{\Sigma }}_r{{\Gamma}_r} \succeq {\bf{\Xi }}_r, \forall  r \in {{\mathcal{R}} _p}, \tag{31c} \label{eq31c}
\end{align}
where the constraint in \eqref{eq31c} stacks the constraints for \eqref{eq26} and \eqref{eq27}. The variables satisfy

\begin{align}
\setcounter{equation}{31}
{\left\{{\begin{array}{*{20}{l}}
			\begin{array}{l}
				{{\bf{\Sigma }}_r} = \\
				\left[ {\begin{array}{*{20}{c}}
						{{\mu _k}{{\bf{I}}_N}\! +\! {\bf{VT}}{{\bf{V}}^H}}&{{\bf{VT}}{{\bf{V}}^H}{{{\bf{\hat h}}}_{a,r}}}\\
						{{\bf{\hat h}}_{a,r}^H{\bf{VT}}{{\bf{V}}^H}}&{ \!-\! {\mu _r}\varepsilon _r^2\! +\! \sigma _{a,r}^2\! +\! {\bf{\hat h}}_{a,r}^H{\bf{VT}}{{\bf{V}}^H}{{{\bf{\hat h}}}_{a,r}},}
				\end{array}} \right],\\
				\forall r \in {\cal R},
			\end{array}\\
		\kern 4pt	{{{\bf{\Xi }}_r} = {\bf{U}}_{a,r}^H\sum\limits_{k \in \mathcal{K}} {{{\bf{W}}_k}} {{\bf{U}}_{a,r}},\forall r \in {\cal R},} \\
		\kern 4pt	{{{\bf{\Sigma }}_{R + 1}} = {{\bf{I}}_N}\Phi _N^{ - 1}\left( {1 - {\kappa ^{1/Q}}} \right)\sigma _p^2 + {\bf{\bar VT}}{{{\bf{\bar V}}}^H},}\\
		\kern 4pt	{{{\bf{\Xi }}_{R + 1}} = \sum\limits_{k \in  \mathcal{K}} {{{\bf{W}}_k}} .}
	\end{array}} \right.}
\label{eq32}\end{align}

Then, the min-max problem \textbf{P6} reduces to a smooth objective function, which is presented in the following theorem.

\textsl{Theorem 1:} The finite min-max problem  \textbf{P6} is replaced by unconstrained minimization of
\begin{align}
\mathop {\min }\limits_{\left\{{{\Gamma }_r}\right\}}  {\phi _p}\left( {{\Gamma _r}} \right),
\label{eq33}\end{align}
where smooth objective function ${\phi _p}\left( {{\Gamma _r}} \right)$ is given by
\begin{align}
{\phi _p}\left( {{\Gamma _r}} \right) = &{p^{ - 1}}{\rm{In}}\sum\limits_{r \in {{{\cal R}}_p}} {{e^{p {{\mathscr{P} }\left( {{\Gamma _r}} \right)}}}}\notag \\&
+ {p^{ - 1}}\sum\limits_{r \in {{{\cal R}}_p}} {{{\boldsymbol{1}}^T}} {e^{p\left[ {{\rm{vec}}\left( {{{\bf{\Xi }}_{r}}} \right){\rm{ - vec}}\left( {{{\bf{\Sigma }}_{r}}} \right){\Gamma _{r}}} \right] - 1}}.
\label{eq34}\end{align}

\begin{proof}
Please refer to Appendix C.
\end{proof}

Due to the equivalence between min-max problem and nonlinear programming problem, the original problem is converted into an unconstrained minimization problem with a smooth objective function ${\phi _p}\left( {{\Gamma _r}} \right)$. This greatly facilitates the numerical solution. To handle optimization problem \eqref{eq33}, we resort to the BFGS algorithm \cite{BFGS-Nezhad}. The BFGS method, as a member of the quasi-Newton methods, is a very general framework that is able to solve unconstrained optimization problem, and has a superior convergence rate.

A rudimentary BFGS algorithm is outlined as follows:

{\bf{Step1}}. Find a starting point ${\{{\Gamma _r}\}^{(1)}}$, set $s:=1$, and pick up convergence tolerance $\epsilon$;

{\bf{Step2}}. Compute the search direction: ${ \{\Delta{\Gamma _r}\}} =-{\bf{X}}\{{\nabla _{\{{\Gamma _r}\}}}{\phi _p}\left( \{{\Gamma _r}\}^{(s)} \right)\}$;

{\bf{Step3}}. Update the dual variables: ${\{{\Gamma _r}\}^{(s+1)}}: = {\{{\Gamma _r}\}^{(s)}} +  {\nu} \{\Delta{\Gamma _r}\}$, where ${\nu} = \arg \mathop {\min }\limits_{{\nu} \ge 0} \{ {\phi _p}({\{{\Gamma _r}\}^{(s)}} + \nu  \{{\Delta\Gamma _r}\})\}$ denotes the step length;

{\bf{Step4}}. Compute iterate variables: ${\boldsymbol{\varrho}}={\nu}{\{\Delta{\Gamma _r}\}}$, and ${{\boldsymbol{\tau}}} ={\nabla _{\{{\Gamma _r}\}}}{\phi _p}\left( \{{\Gamma _r}\}^{(s+1)} \right)- {\nabla _{\{{\Gamma _r}\}}}{\phi _p}\left( \{{\Gamma _r}\}^{(s)} \right)$;

{\bf{Step5}}. Update ${{\bf{X}}^{(s + 1)}}: = \left( {{\bf{I}} - \delta {\boldsymbol{\varrho}}{{\boldsymbol{\tau}}^T}} \right){{\bf{X}}^{(s)}}\left( {{\bf{I}} - \delta {\boldsymbol{\tau}}{{\boldsymbol{\varrho}}^T}} \right) + \delta {\boldsymbol{\varrho}}{{\boldsymbol{\varrho}}^T}$, $\delta  = 1/{{\boldsymbol{\varrho}}^T}{\boldsymbol{\tau}}$;

{\bf{Step6}}. If $\| {\boldsymbol{\varrho}} \|_2^2 < \epsilon $, terminate the algorithm and output;

{\bf{Step7}}. Otherwise, $s:=s+1$, go back to Step 2.

\begin{table}[t]
\begin{center}
\begin{tabular}{llr}
\hline
\textbf{Algorithm 1 }Numerical algorithm for problem \textbf{P5}\\
\hline
\textbf{Initialization:} Pick up ${\boldsymbol{\varrho}}:={\boldsymbol{1}}, {{\boldsymbol{\tau }}}:={\boldsymbol{0}}, {{\bf{X}}}: = {\bf{I}}$, and  \\ \kern 10pt
tolerance  $\epsilon>0$; \\
1. \textbf{Loop}\\
2. \kern 10pt	Solve \textbf{P4} based on SDP to obtain $\left({\left\{ {{{\bf{W}}_k}} \right\}}_{k \in {{\cal K}}},{\bf{T}}, {\boldsymbol{\mu}}\right)$;\\
3. \kern 10pt 	Determine $\{{\bf{\Sigma }}_r\}_{r \in {{{\cal {R}}_{p}}}}$ and $\{{\bf{\Xi }}_r\}_{r \in {{{\cal {R}}_{p}}}}$ by $\left({\left\{ {{{\bf{W}}_k}} \right\}}_{k \in {{\cal K}}},{\bf{T}}, {\boldsymbol{\mu}}\right)$;\\
4. \kern 10pt   Update ${\boldsymbol{\varrho}}$, ${{\boldsymbol{\tau }}}$, and ${{\bf{X}}}$ following BFGS algorithm; \\
5. \kern 10pt   Some stopping criterion is satisfied; \\
6. \textbf{End loop}  \\
\textbf{Output:} Get the optimal solution $\left({\left\{ {{{\bf{W}}^\star_k}} \right\}}_{k \in {{\cal K}}},{\bf{T}}^\star, {\boldsymbol{\mu}}^\star,\left\{\Gamma _r^\star\right\}_{r \in {{{\cal {R}}_p}}}\right)$.\\
\hline
\end{tabular}
\label{tab3}
\end{center}
\end{table}
Algorithm 1 summarizes the exponential penalty method associated with BFGS procedure to solve \textbf{P5}.

Finally, we summarize the procedure of the two-stage algorithm for the transmit power minimization problem. We jointly design transmit information and AN beamforming with fixed SINR tolerance values in the first stage. Next, using the transmit information and AN beamforming obtained in the first stage, we determine the minimum maximum Eve achievable rate values in the second stage. In particular, this two-stage algorithm is done not only to achieve the minimum transmit power on condition of providing QoS assurance for each LU, but to beat the target of maximum secrecy rate. With each stage being solved analytically, the two-stage algorithm can be efficiently performed.

\subsection{Optimality for Received Weight Vector}

We assume LUs can obtain the directional information of transmitter by the handshaking signals, while any information of the jamming signals is unavailable. In such circumstances, we adopt the MVDR method {\cite{Optimum-array-processing-Van}} to design the received weight vector for suppressing the jamming from AEs. The propose of MVDR method is to minimize the interference-plus-noise power subject to maintaining a distortionless response to the desired direction. The anti-jamming processing at LU $k$ is then formulated as follows:
\begin{align}
	\textbf{P7:}\kern 4pt	&\mathop \text{min}\limits_{{\bf{v}}_k} \kern 2pt  {\bf{v}}_k^H{{\bf{R}}_y}{\bf{v}}_k \tag{35a} \label{eq35a}\\
	& {\rm{s.t.}} \kern 2pt {\bf{a}}_{u,k}^H{\bf{v}}_k= 1. \tag{35b} \label{eq35b}
\end{align}
The solution of \textbf{P7} is given by
\begin{eqnarray}
	\setcounter{equation}{36}
\begin{aligned}[b]
{{\bf{v}}^ \star_k } = {\bf{R}}_y^{ - 1}{{\bf{a}}_{u,k}}{\left({{\bf{a}}_{u,k}^H{\bf{R}}_y^{ - 1}{{\bf{a}}_{u,k}}}\right) ^{ - 1}},
\end{aligned}
\label{eq39}\end{eqnarray}
where the covariance matrix ${{\bf{R}}_y}$ is hard to obtain in practice. Hence, we usually use the sample covariance matrix of each antenna instead of the covariance matrix, i.e., the sample covariance matrix ${{{\bf{\tilde R}}}_y} = \frac{1}{L}\sum\limits_{i = 1}^L {{{\bf{y}}_i}{\bf{y}}_i^H}$ with $\{ {{\bf{y}}_i}\} _1^L$ being the received signal snapshots, and $L$ being the length of snapshots. 

\textsl{Remark 4:} Based on the DoF of array, it is requested that the antenna number of received array more than that of AEs, i.e., $M>R$,  to effectively suppressed jamming from AEs. Due to the small wavelengths in mmWave band, the eligible antenna space can be very small, and hence large-scale arrays are feasible in both transmitter and receiver sides in mmWave communication systems.

\section{Simulation Results}
In this part, different simulation scenarios are provided to elaborate the secrecy performance of our proposed scheme. We consider a LoS mmWave communication system with carrier frequency ${f_c} = 30$ {\rm{GHz}}. The system consists of an $N$-antenna transmitter, and $K$ $M$-antenna LUs with direction ${\theta _{u,k}}, \forall k \in \mathcal{K}$. Unless specified otherwise, we set the LUs' number as $K=2$ and the directions as ${\theta _{u,1}}=-35^\circ$, ${\theta _{u,2}}=15^\circ$. To focus on the beamforming characteristics, the reference distances are set as 1000 meters. Hence, the assumption of far-field transmission holds. The power constraints for per-antenna are set as equal, i.e. $P_n = P_\text{Tol}/N$, $\forall n$, where $P_\text{Tol}$ is the total transmit power at the transmitter. The per-antenna power budget for the transmitter is fixed to $P_n=30$ dBm, $\forall n \in \mathcal{N}$. The antenna spacing of the ULA is $d_t=d_l = c/2f_c$ to avoid creating grating lobes. For simplicity, all the background thermal noise variance is assumed to be identical, i.e., $10\lg (\sigma _{u,k}^2) =10\lg (\sigma _{u}^2)=-100$ \text{dBm}, $\forall k$. We assume that the prescribed minimum received SNRs at each of the LU are equal, i.e., ${\zeta}={\zeta _k}$, $\forall k$. On the other hand, different types of Eves exist in the system. The multipath fading coefficients between transmitter and PEs are modeled as Rayleigh random variables intercepting the confidential messages with number $Q=2$. Besides, there are $R=3$ AEs intercepting the confidential messages and emitting strong jamming signals, whose directions are  ${\theta _{a,1}} =  - 60^\circ$, ${\theta _{a,2}} =  3^\circ$, and ${\theta _{a,3}} =  60^\circ$. In practice, the channel state of Eves is better than that of LUs. Therefore, the channel noise power of Eve is set as $10\lg (\sigma _{p}^2) = 10\lg (\sigma _{a}^2)= -120$ \text{dBm}. The jammer-to-noise ratios (JNRs) from AEs to LUs are set as $30$ {\rm{dB}}. The probability parameter is set as $\kappa=0.95$ for providing secure communication. To facilitate the statement in the sequel, we normalize the uncertainty levels of AE as $\chi=\varepsilon _r^2/{ \| {{{\bf{h}}_{a,r}}} \|_2^2}$, $\forall r$. Additionally, we also compare the performance of our proposed method with the existing method. We choose the main-lobe-integration (MLI)-robust method \cite{Robust-Secure-Transmission-Shu} as the benchmark scheme since both our design and the benchmark scheme use robust beamforming design and are suitable for multiple targets. Following the electromagnetic wave propagation path loss model in the free space \cite{Digital-communications-Proakis}, the path loss factor $\rho \left( r \right)$ is determined by
\begin{eqnarray}
\begin{aligned}[b]
{\rm{Lfs}}(\text{dB}) &= - 20{\lg }[\rho (r)]\\&
= 32.5 + 20\lg[{f_c}(\text{MHz})] + 20\lg[r(\text{Km})],
\end{aligned}
\label{eq37}\end{eqnarray}
where $f_c$ is the transmit carrier frequency in megahertz (MHz), and $r$ is the range in kilometer (Km).
\begin{figure}[tb]
	\begin{center}
		\includegraphics[width=1\columnwidth]{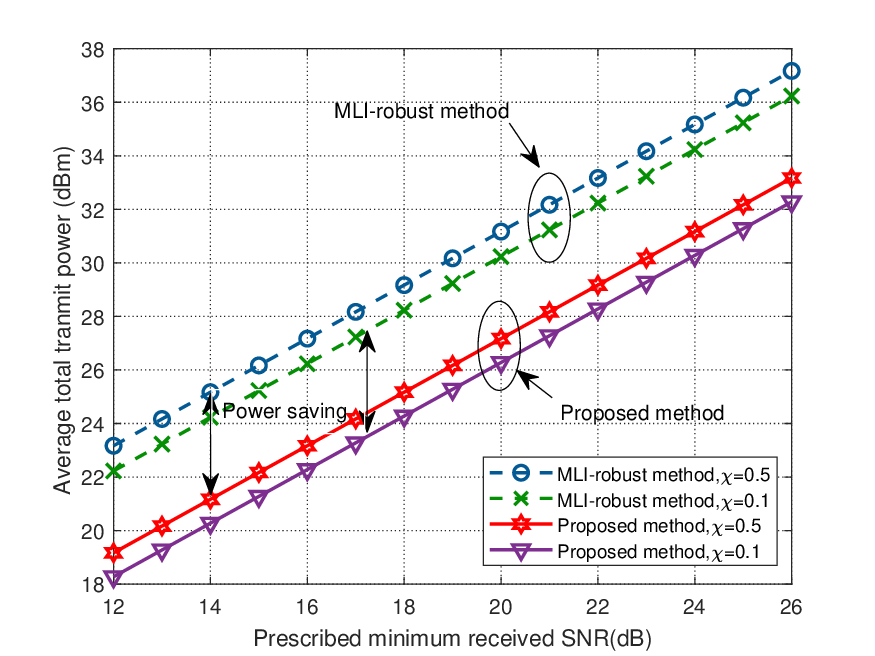}
	\end{center}
	\caption{The average total transmit power versus the prescribed minimum received SNR, for different AE CSI uncertainty levels and the MLI-robust method.}
	\label{fig3}
\end{figure}

In the first scenario, the effect of the prescribed minimum received SNR at each of the LU on average total transmit power is investigated for different AE CSI uncertainty levels $\chi=0.1, 0.5$, and the MLI-robust method, when transmit antenna number is set as $N=20$. A general observation in Fig. \ref{fig3} is that the average total transmit power consumed by the proposed scheme is a monotonically increasing function of the prescribed minimum received SNR, which is attributed to the fact that more transmit power need to be allocated to information transmission for providing the QoS assurance when the minimum required received SNR becomes more stringent. With increasing AE uncertainty levels, it has to consume more transmit power to generate AN to prevent interception of AEs. Besides, we compare the performance of our proposed scheme with that of the MLI-robust method. For a fair comparison, we pick up the same prescribed minimum received SNR and the maximum received SINR tolerance. In particular, it can be observed that the proposed scheme consumes less transmit power compared with the MLI-robust method. Indeed, our proposed scheme fully utilizes the CSI of all communication links and jointly optimizes the space spanned by AN for performing less transmit power consumption. 
\begin{figure}[tb]
	\begin{center}
		\includegraphics[width=1\columnwidth]{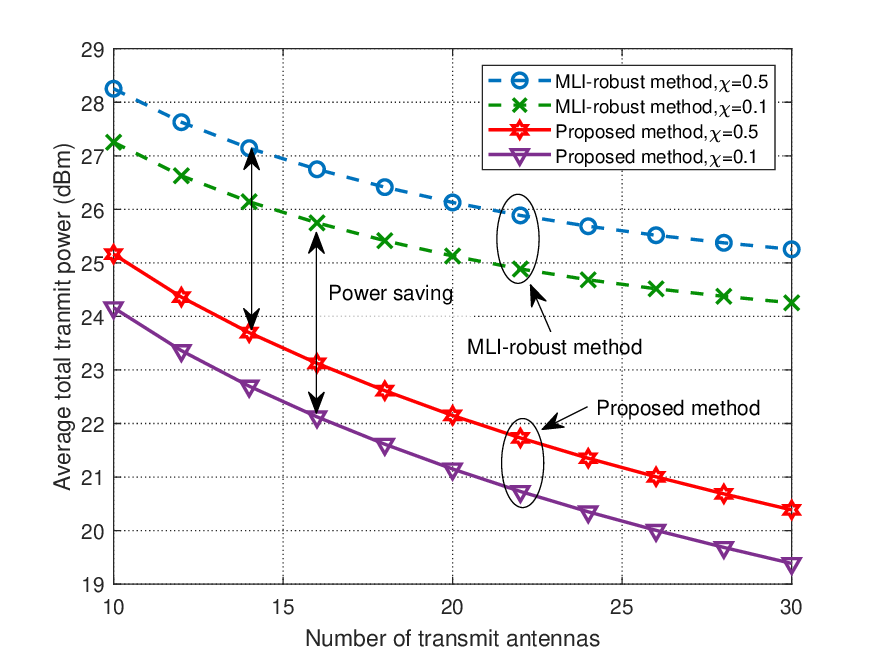}
	\end{center}
	\caption{The average total transmit power versus the number of transmit antennas, for different AE CSI uncertainty levels and the MLI-robust method.}
	\label{fig4}
\end{figure}

Figure \ref{fig4} presents the average total transmit power consumption versus the number of transmit antennas, for different AE CSI uncertainty levels $\chi=0.1, 0.5$, and the MLI-robust method. The prescribed minimum received SNR of LU is set as ${\zeta}=15$ dB. As seen, the consumption of the total transmit power reduces as the number of antennas increases. This is because more transmit antenna enhances the array’s capability in array signal processing, and thus reduces the power consumption. In particular, equipped with more antennas, the transmitter is capacity of more efficient beam and AN interference designs. It can also be observed that our proposed scheme achieves more power savings compared with the MLI-robust method for the joint design of information and AN beamforming. On the other hand, the MLI-robust method consumes more transmit power since it can not fully apply the potential DoF. Again, with larger uncertainty levels, the average total transmit power consumption increases.
\begin{figure}[tb]
	\begin{center}
		\includegraphics[width=1\columnwidth]{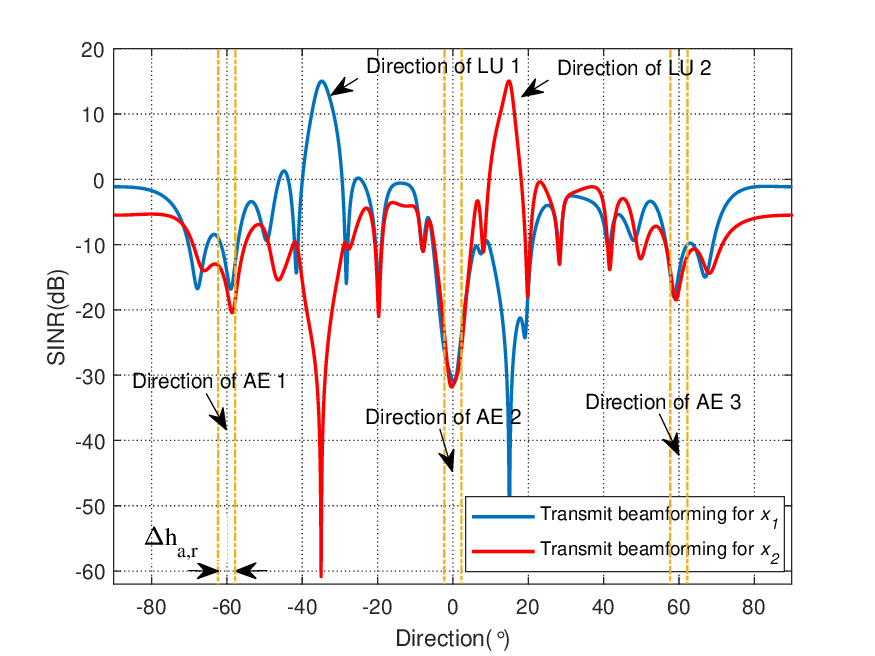}
	\end{center}
	\caption{The beampattern of transmitter versus direction of the proposed method.}
	\label{fig5}
\end{figure}

Based on the discussion in Section II, we know that the multiple confidential messages need to simultaneously transmit toward corresponding LUs by our proposed beamforming and AN joint design. Typically, in the case of $N=20$, we investigate the beampattern of transmitter versus direction dimension in Fig. \ref{fig5}. Sure enough, two sharp SINR peaks are formed in the directions of LUs. The signals along with the directions of LUs are fully compliant with the predefined requirements, 15 dB. These guarantee a reliable transmission from the transmitter to LUs. Additionally, the SINRs are so poor along other directions for the weak message power leakage and strong AN interference, especially along the directions of AEs.
\begin{figure}[tb]
	\begin{center}
		\includegraphics[width=1\columnwidth]{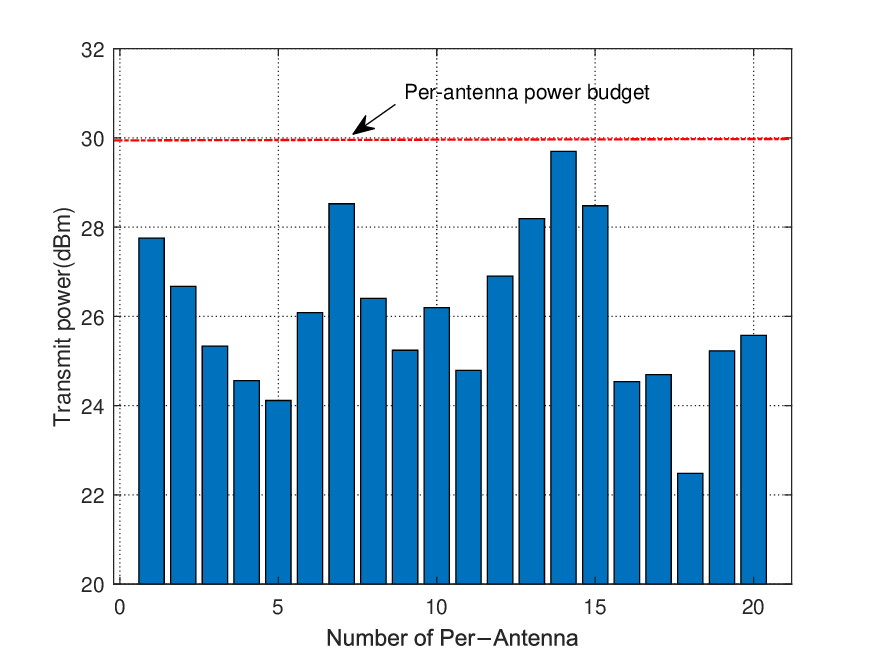}
	\end{center}
	\caption{The average transmit power versus the index of transmit antennas.}
	\label{fig6}
\end{figure}

To show whether the per-antenna maximum transmit power constraint can be satisfied, we list the per-antenna transmit power consumption versus the index of transmit antennas in Fig. \ref{fig6}. We set the per-antenna power budget as $P_n=30$ dBm, $\forall n$. It can be found that the transmit power meets the per-antenna power constraints. As mentioned earlier, per-antenna power constraints are more practically relevant since each antenna usually has its own power amplifier.
\begin{figure}[tb]
	\begin{centering}
		\subfloat[]{\begin{centering}
				\includegraphics[width=1\columnwidth]{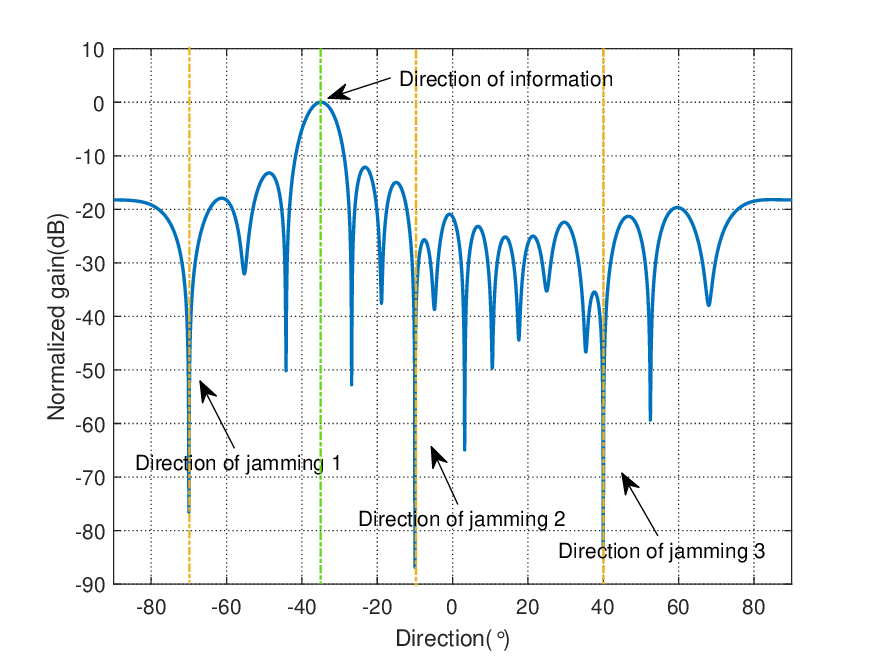}
				\par\end{centering}
		}\\
		\subfloat[]{\begin{centering}
				\includegraphics[width=1\columnwidth]{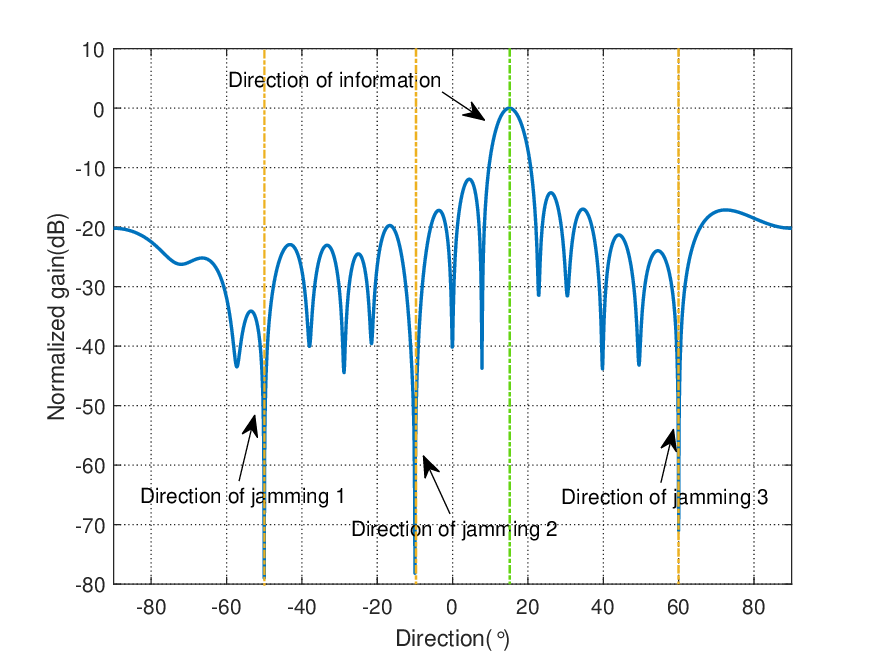}
				\par\end{centering}
		}
		\par\end{centering}
	\caption{The beampattern of the LU versus direction of the proposed method for (a) LU 1, (b) LU 2.}
	\label{fig7}\end{figure}

In next simulation, we fix the jamming directions as ${\theta _{j,1,1}}=-70^\circ$, ${\theta _{j,1,2}}=-10^\circ$, ${\theta _{j,1,3}}=40^\circ$, ${\theta _{j,2,1}}=-50^\circ$, ${\theta _{j,2,2}}=-10^\circ$, ${\theta _{j,2,3}}=60^\circ$, and the number of received antennas is set as $M=16$. Fig. \ref{fig7} illustrates the received beampattern of the LU 1 and 2, respectively, versus direction dimension of our proposed method. It can be seen that three deep nulls are formed along the directions of the jamming signals, as well as the signal gain along the transmit direction is zero. The results indicate the jamming signals emitted by AEs are able to be effectively suppressed, while maintaining a distortionless response to the signals from the transmit direction.
\begin{figure}[tb]
	\begin{center}
		\includegraphics[width=1\columnwidth]{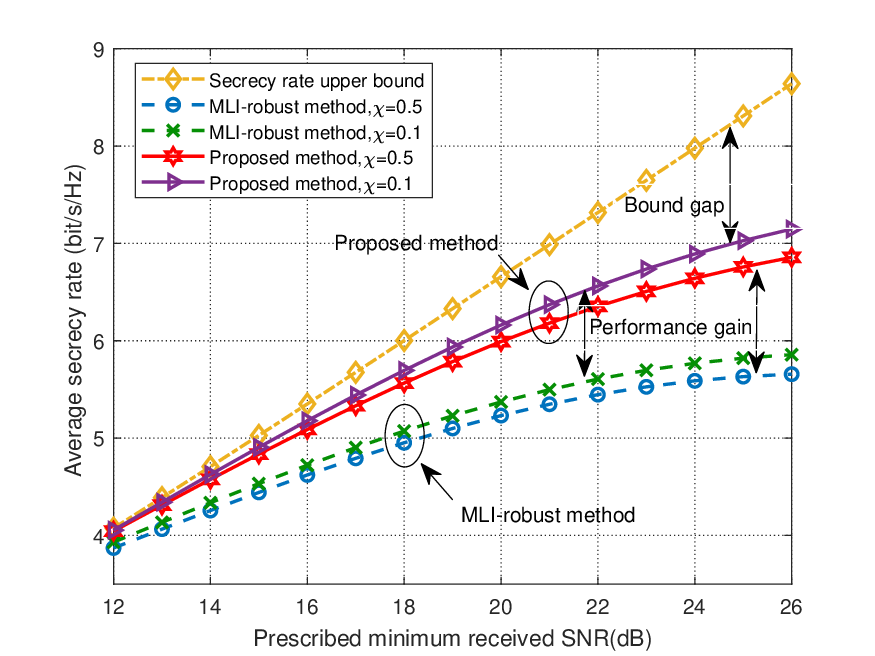}
	\end{center}
	\caption{The average system secrecy capacity (bit/s/Hz) versus the prescribed minimum received SNR, for different AE CSI uncertainty levels and the MLI-robust method.}
	\label{fig8}
\end{figure}

Finally, we further plot the average system secrecy rate versus the prescribed minimum received SNR in Fig. \ref{fig8} for different AE CSI uncertainty levels $\chi=0.1, 0.5$, and the MLI-robust method, with $N=20$, $M=16$. For a fair comparison of the secrecy performance, we pick up the same prescribed minimum received SNR and the total transmit power. Moreover, we also show the secrecy rate upper bound, i.e., the maximum LU achievable rate under no Eve existence in the system. As can be seen from Fig. \ref{fig8}, the average system secrecy capacity increases with prescribed minimum received SNR since the LU achievable rate is increasing when more prescribed minimum received SNR is required. As expected, our proposed scheme achieves a higher secrecy capacity in the high prescribed minimum received SNR regime compared with the MLI-robust method. The superior performance in secrecy rate stems from the transmit power saving. On the other hand, the gap between the average system secrecy rate and the secrecy rate upper bound is growing in a higher minimum received SNR regime since per-antenna maximum transmit power limits the secrecy performance.

\section{Conclusion}
In this paper, we considered the PHY security problem for MU-MIMO mmWave wireless communication in the presence of hybrid Eves, under which many conventional methods may fail to offer satisfactory secrecy performance. To tackle this, we proposed an AN-aided robust multi-beam scheme to achieve secure and reliable mmWave wireless transmission. In particular, we minimized the total transmission power by jointly designing the information and AN beamforming, subject to constraints on received SNR for each LU, tolerant SINR for Eves, and per-antenna power for transmitter. Additionally, by means of MVDR method to design the received weight vector, the LUs were able to effectively suppress the jamming signals from AEs. Considering the intractability of the total transmit power minimization problem, we first reformulated the non-convex constraints as tractable convex constraints. Then, we got the optimal solution of the reformulated problem based on SDP. Moreover, we formulated a min-max problem to enhance the system secrecy performance. To obtain the global optimal solution, we resorted to the exponential penalty method associated with the BFGS algorithm. In the end, the superior performance in terms of the total transmit power savings and the security has been illustrated via extensive numerical results.

\appendices
\section{Proof of Lemma 2}
Considering $Q$ independent channels of PEs, the left-hand side in constraint \eqref{eq21d} for wiretapping the confidential messages is given by
\begin{eqnarray}
&\Pr \left\{ {\mathop {\max }\limits_{q \in  {\cal Q}} \frac{{\sum\limits_{k \in  {\cal K}} {{\rm{Tr}}} \left( {{{\bf{H}}_{p,q}}{{\bf{W}}_k}} \right)}}{{{\rm{Tr}}\left( {{{\bf{H}}_{p,q}}{\bf{\bar VT}}{{{\bf{\bar V}}}^H}} \right) + \sigma _{p}^2}} \le {\Gamma _p}} \right\}\notag\\&
= \prod\limits_{q \in  {\cal Q}} {\Pr \left\{ {\frac{{\sum\limits_{k \in  {\cal K}} {{\rm{Tr}}} \left( {{{\bf{H}}_{p,q}}{{\bf{W}}_k}} \right)}}{{{\rm{Tr}}\left( {{{\bf{H}}_{p,q}}{\bf{\bar VT}}{{{\bf{\bar V}}}^H}} \right) + \sigma _{p}^2}} \le {\Gamma _p}}\right\}}.
\label{eq38}\end{eqnarray}

Therefore, by some mathematical manipulations, constraint \eqref{eq21d} is equivalently expressed as follows:
\begin{align}
&\Pr \left\{ {\mathop {\max }\limits_{q \in {\cal Q}} \frac{{\sum\limits_{k \in {\cal K}} {{\rm{Tr}}} \left( {{{\bf{H}}_{p,q}}{{\bf{W}}_k}} \right)}}{{{\rm{Tr}}\left( {{{\bf{H}}_{p,q}}{\bf{\bar VT}}{{{\bf{\bar V}}}^H}} \right) + \sigma _{p}^2}} \le {\Gamma _{p}}} \right\} \ge \kappa \label{eq39}\\&
\Leftrightarrow \Pr \left\{ {{\rm{Tr}}\left( {{{\bf{H}}_{p}}{\bf{{P}}}} \right) \le {\Gamma _p}\sigma _{p}^2} \right\} \ge {\kappa ^{1/Q}},
\label{eq40} \end{align}
where ${\bf{{P}}} = \sum\limits_{k \in {\cal K}} {{{\bf{W}}_k}}  - {\Gamma _p}{\bf{\bar VT}}{{{\bf{\bar V}}}^H}$. Due to the equivalent random channels of PEs following i.i.d. random variables, we drop the index of PE channels.

Next, the probabilistic constraint is replaced with its upper bound, i.e., 
\begin{eqnarray}
\begin{aligned}[b]
{\mathop{\rm Tr}\nolimits} \left( {{{\bf{H}}_p}{\bf{P}}} \right) & \overset{{(a)}}\le \sum\limits_{i = 1}^N {{\lambda _i}} \left( {{{\bf{H}}_p}} \right){\lambda _i}({\bf{P}})\\&
\overset{{(b)}}= {\lambda _{\max }}\left( {{{\bf{H}}_p}} \right){\lambda _{\max }}({\bf{P}})\\&
\overset{{(c)}}= {\mathop{\rm Tr}\nolimits} \left( {{{\bf{H}}_p}} \right){\lambda _{\max }}({\bf{P}}),
\end{aligned}
\label{eq41}\end{eqnarray}
where ${{\lambda _i}}\left( \cdot  \right)$ denotes the  $i$ eigenvalue of matrix and its orders are arranged ${\lambda _{\max }}\left(  \cdot  \right) = {\lambda _1}\left(  \cdot  \right) \ge {\lambda _i}\left(  \cdot  \right) \ge ... \ge {\lambda _N}\left(  \cdot  \right) = {\lambda _{\min }}\left(  \cdot  \right)$. In addition, the inequality $(a)$ is due to the trace inequality for positive Hermitian matrices \cite{Inequalities-Marshall}, while the equalities $(b)$ and $(c)$ can hold since ${{{\bf{H}}_p}}$ is a rank-one positive semidefinite matrix. Utilizing \eqref{eq40} and \eqref{eq41}, we obtain the inequality as follows:
\begin{align}
\Pr \left\{ {{\rm{Tr}}\left( {{{\bf{H}}_p}{\bf{P}}} \right) \le {\Gamma _p}\sigma _p^2} \right\} \ge \Pr \left\{ {{\rm{Tr}}\left( {{{\bf{H}}_p}} \right){\lambda _{\max }}\left( {\bf{P}} \right) \le {\Gamma _p}\sigma _p^2} \right\}.
\label{eq42}\end{align}
According to the previous discussion, we have
\begin{eqnarray}
\begin{aligned}[b]
&\Pr \left\{ {\mathop {\max }\limits_{q \in  {\cal Q}} \frac{{\sum\limits_{k \in  {\cal K}} {{\rm{Tr}}} \left( {{{\bf{H}}_{p,q}}{{\bf{W}}_k}} \right)}}{{{\rm{Tr}}\left( {{{\bf{H}}_{p,q}}{\bf{\bar VT}}{{{\bf{\bar V}}}^H}} \right) + \sigma _{p}^2}} \le {\Gamma _p}} \right\}\\&
\ge \Pr \left\{ {{\rm{Tr}}\left( {{{\bf{H}}_p}} \right){\lambda _{\max }}\left( {\bf{P}} \right) \le {\Gamma _p}\sigma _p^2} \right\} \ge {\kappa ^{1/Q}}\\&
\overset{{(d)}} \Leftrightarrow \Pr \left\{ {\frac{{{\lambda _{\max }}\left( {\bf{P}} \right)}}{{{\Gamma _p}\sigma _p^2}} \le \frac{1}{{{\rm{Tr}}\left( {{{\bf{H}}_p}} \right)}}} \right\} \ge {\kappa ^{1/Q}}\\&
\overset{{(e)}} \Leftrightarrow \Pr \left\{ {\frac{1}{{{\rm{Tr}}\left( {{{\bf{H}}_p}} \right)}} \le \frac{{{\lambda _{\max }}\left( {\bf{P}} \right)}}{{{\Gamma _p}\sigma _p^2}}} \right\} \le 1 - {\kappa ^{1/Q}}\\&
\overset{{(f)}} \Leftrightarrow {\lambda _{\max }}\left( {\bf{P}} \right) \le \Phi _N^{ - 1}\left( {1 - {\kappa ^{1/Q}}} \right){\Gamma _p}\sigma _p^2\\&
\Leftrightarrow {\bf{P}} \preceq {{\bf{I}}_N}\left[ {\Phi _N^{ - 1}\left( {1 - {\kappa ^{1/Q}}} \right){\Gamma _p}\sigma _p^2} \right],
\end{aligned}
\label{eq43}\end{eqnarray}
where equivalent transformation $(d)$ and $(e)$ hold for the positive definiteness of matrix ${{\bf{H}}_p}$ and the basic property of probability, respectively. Equivalent transformation $(f)$ is obtained similarly to the steps of \cite[Lemma 1]{Joint-Information-Nguyen}. $\Phi _N^{ - 1}\left(  \cdot  \right)$ represents the inverse c.d.f. of an inverse central chi-square random variable with $2N$ DoF. Actually, the inverse function of the inverse central chi-square c.d.f. can be evaluated directly or be stored in a lookup table in a practical implementation. Thus, Lemma 2 follows.

\section{Proof of Lemma 3}
It can be observed that the relaxed version of problem \textbf{P4} is jointly convex with respect to the optimization variables and satisfies Slater’s constraint qualification. Therefore, the Karush-Kuhn-Tucker (KKT) conditions are sufficient conditions to solve the relaxed problem \cite{Convex-optimization-Boyd}. The Lagrangian function of problem \textbf{P4}, denoted by ${\cal L}\left( {\left\{ {{{\bf{W}}_k}} \right\},{\bf{T}},{\boldsymbol{\mu }},\left\{ {{{\bf{D}}_k}} \right\},\left\{ {{{\bf{B}}_r}} \right\},\left\{ {{{\bf{G}}_k}} \right\},\left\{ {{\alpha _k}} \right\},\left\{ {{\beta _n}} \right\}} \right)$, can be derived as follows:
\begin{align}
&{\cal L}\left( {\left\{ {{{\bf{W}}_k}} \right\},{\bf{T}},{\boldsymbol{\mu }},\left\{ {{{\bf{D}}_k}} \right\},\left\{ {{{\bf{B}}_r}} \right\},\left\{ {{{\bf{G}}_k}} \right\},\left\{ {{\alpha _k}} \right\},\left\{ {{\beta _n}} \right\}} \right)\notag\\&
= \sum\limits_{k \in {\cal K}} {{\rm{Tr}}\left( {{{\bf{W}}_k}} \right)} - \sum\limits_{r \in \cal R} {{\rm{Tr}}\left\{ {{{\bf{B}}_r}{{\bf{S}}_r}\left( {{{\bf{W}}_k},{\bf{T}},{\mu _r}} \right)} \right\}} \notag\\&
\kern 8pt + \sum\limits_{n \in {\cal N}} {{\beta _n}\sum\limits_{k \in {\cal K}} {{\rm{Tr}}\left( {{{\bf{E}}^{(n)}}{{\bf{W}}_k}} \right)} }  + \sum\limits_{k \in {\cal K}} {{\rm{Tr}}\left( {{{\bf{G}}_k}{{\bf{W}}_k}} \right)}\notag \\&
\kern 8pt - \sum\limits_{k \in {\cal K}} {{\alpha _k}{\rm{Tr}}\left( {{{\bf{H}}_{u,k}}{{\bf{W}}_k}} \right)}   - \sum\limits_{k \in {\cal K}} {{\rm{Tr}}\left( {{{\bf{D}}_k}{{\bf{W}}_k}} \right)}  + {{O}},
\label{eq44}\end{align}
where ${{O}}$ is the collection of terms that are not relevant. In \eqref{eq44}, ${\alpha _k}\ge 0$, ${k \in  {\cal K}}$, are the Lagrange multiplier of the prescribed minimum received SNR in constraints \eqref{eq29b}. ${{\bf{B}}_r}\succeq  \boldsymbol{0}$, ${r \in  {\cal R}}$, and ${{\bf{G}}_k}\succeq  \boldsymbol{0}$, ${k \in  {\cal K}}$, indicate the Lagrange multiplier matrices of the corresponding to constraints \eqref{eq29c} and \eqref{eq29d} on the Eve tolerance, respectively. ${\beta _n}\ge 0$, ${n \in  {\cal N}}$, are the Lagrange multiplier corresponding to constraint \eqref{eq29e} on the per-antenna maximum transmit power. Matrices ${{\bf{D}}_k}\succeq  \boldsymbol{0}$, ${k \in  {\cal K}}$, are the Lagrange multiplier matrices for the positive semi-definite constraint on matrix ${{{\bf{W}}_k}}$, ${k \in  {\cal K}}$, in constraint \eqref{eq29g}. Then, the KKT conditions with respect to  ${{{\bf{W}}_k^\star}}$, ${k \in  {\cal K}}$, are given by
\begin{subequations}
	\begin{numcases}{}
		\left\{ {{{\bf{D}}_k^\star}} \right\}\succeq  \boldsymbol{0},\forall k \in {\cal K},\label{eq45a}
		\\
		\left\{ {{{\bf{B}}_r^\star}} \right\}\succeq  \boldsymbol{0},\forall r \in {\cal R},\label{eq45b}
		\\
		\left\{ {{{\bf{G}}_k^\star}} \right\}\succeq  \boldsymbol{0}, \forall k \in {\cal K},\label{eq45c}
		\\
		\left\{ {{\alpha _k^\star}} \right\} \ge 0,\forall k \in {\cal K},\label{eq45d}
		\\
		\left\{ {{\beta _n^\star}} \right\}\ge 0,\forall n \in {\cal N},\label{eq45e}
		\\
		{{{\bf{D}}_k^\star}{{\bf{W}}_k^\star}}=\boldsymbol{0},\forall k \in {\cal K},\label{eq45f}
		\\
		{\nabla _{{{\bf{W}}_k}}}{\cal L}=\boldsymbol{0},\forall k \in {\cal K},\label{eq45g}
	\end{numcases}
\end{subequations}
where $\left(\left\{ {{{\bf{D}}_k^\star}} \right\},\left\{ {{{\bf{B}}_r^\star}} \right\},\left\{ {{{\bf{G}}_k^\star}} \right\},\left\{ {{\alpha _k^\star}} \right\},\left\{ {{\beta _n^\star}} \right\}\right)$ denote the optimal Lagrange multipliers of the dual problem \textbf{P4}. Additionally, $\left\{ {{\alpha _k^\star}} \right\} \ge 0,\left\{ {{\beta _n^\star}} \right\}\ge 0$ should hold for ${\zeta _k}>0$ and $P_n>0$. \eqref{eq45f} is the complementary slackness condition and should be satisfied, i.e., the columns of ${{\bf{W}}_k^\star}$ should lie in the null space of ${{\bf{D}}_k^\star}$ for ${{\bf{W}}_k^\star} \ne \boldsymbol{0}$, ${k \in  {\cal K}}$. Then, the KKT condition in \eqref{eq45g} can be further expanded as follows:
\begin{align}
{{\bf{I}}_N} + {{\bf{G}}_k^{ \star }} + \sum\limits_{n \in {\cal N}} {{\beta _n^{ \star }}{{\bf{E}}^{(n)}}}  + \sum\limits_{r \in {\cal R}} {{{\bf{U}}_{a,r}}{{\bf{B}}_r^{ \star }}{\bf{U}}_{a,r}^H}  = {{\bf{D}}_k^{ \star }} + {\alpha _k^{ \star }}{{\bf{H}}_{u,k}}.
\label{eq46}\end{align}
To simplify the notations, we define
\begin{align}
{{\boldsymbol{\Lambda}}_k^{ \star }} \buildrel \Delta \over = {{\bf{I}}_N} + {{\bf{G}}_k^{ \star }} + \sum\limits_{n \in {\cal N}} {{\beta _n^{ \star }}{{\bf{E}}^{(n)}}}  + \sum\limits_{r \in {\cal R}} {{{\bf{U}}_{a,r}}{{\bf{B}}_r^{ \star }}{\bf{U}}_{a,r}^H} .
\label{eq47}\end{align}
Therefore, we have
\begin{align}
{{\boldsymbol{\Lambda}}_k^{ \star }} = {{\bf{D}}_k^{ \star }}+ {\alpha _k^{ \star }}{{\bf{H}}_{u,k}}.
\label{eq48}\end{align}
Then, both sides of \eqref{eq48} are post-multiplying by ${{\bf{W}}_k^{ \star }}$, yielding
\begin{align}
{{\boldsymbol{\Lambda}}_k^{ \star }}{{\bf{W}}_k^{ \star }}& = \left({{\bf{D}}_k^{ \star }}+ {\alpha _k^{ \star }}{{\bf{H}}_{u,k}}\right){{\bf{W}}_k^{ \star }}\notag\\&
\overset{{(a)}}=  {\alpha _k^{ \star }}{{\bf{H}}_{u,k}}{{\bf{W}}_k^{ \star }},
\label{eq49}\end{align}
where the equality $(a)$ is due to the complementary slackness condition in \eqref{eq45f}. 

We note that ${{\boldsymbol{\Lambda}}_k^{ \star }}$ is a positive definite matrix, i.e., ${{\boldsymbol{\Lambda}}_k^{ \star }}\succeq  \boldsymbol{0}$. Consequently, by utilizing \eqref{eq49} and a basic rank inequality, we obtain
\begin{align}
{\rm{Rank}}\left({{\bf{W}}_k^{ \star }}\right)&={\rm{Rank}}\left({{\boldsymbol{\Lambda}}_k^{ \star }}{{\bf{W}}_k^{ \star }}\right) \notag\\&
={\rm{Rank}}\left({\alpha _k^{ \star }}{{\bf{H}}_{u,k}}{{\bf{W}}_k^{ \star }}\right)\notag\\&
=\min \left\{{\rm{Rank}}\left({{\bf{W}}_k^{ \star }}\right),{\rm{Rank}}\left({\alpha _k^{ \star }}{{\bf{H}}_{u,k}}\right)\right\}.
\label{eq50}\end{align}
It is obvious that ${\rm{Rank}}\left({{\bf{H}}_{u,k}}\right)=1$, and thus ${\rm{Rank}}\left({\alpha _k^{ \star }}{{\bf{H}}_{u,k}}\right)=1$ for ${\alpha _k^{ \star }}\ge 0$. In addition, due to the received SNR conditions for LUs, ${{\bf{W}}_k^{ \star }} \ne \boldsymbol{0}$. Therefore, ${\rm{Rank}}\left({{\bf{W}}_{k}}\right)=1$, $\forall k \in {{\cal K}}$.
Thus, Lemma 3 follows.

\section{Proof of Theorem 1}
According to \cite[Theorem 1]{exponential-penalty-Li}, the original constrained optimization problem \textbf{P6} is transformed as the following smooth unconstrained optimization problem: 
\begin{align}
	{\Psi _p}(\omega, \left\{{\Gamma _r}\right\}) = & \omega  + {p^{ - 1}}\sum\limits_{r \in {{\mathcal{R}} _p}} e^{ {p\left[ { \mathscr{P} \left( {{\Gamma _r}} \right) - \omega } \right] - 1} } \notag\\&
	+ {p^{ - 1}}\sum\limits_{r \in {{\mathcal{R}} _p}} \boldsymbol{1}^T {e ^{ { {p\left[ {\rm{vec}\left({{\bf{\Xi }}_r}\right) - \rm{vec}\left({{\bf{\Sigma }}_r}\right){\Gamma _r}}\right] - 1}}}}.
	\label{eq51}\end{align}
Since ${\Psi _p}(\omega, \left\{{\Gamma _r}\right\})$ is a convex function of variable $\omega$ for any $\left\{{\Gamma _r}\right\}_{r \in {{\mathcal{R}} _p}}$, the optimal condition with respect to $\omega$ to be a solution of nonlinear programming problem need to satisfy the following condition:
\begin{align}
	\partial {\Psi _p}(\omega ,\left\{{\Gamma _r}\right\})/\partial \omega  = 1 - \sum\limits_{r \in {{{\cal R}}_p}} {{e^{p\left[ {{ \mathscr{P}}\left( {{\Gamma _r}} \right) - \omega } \right] - 1}}}  = 0.
	\label{eq52}\end{align}
Then, we can obtain the following relationship between the optimal variables ${\omega ^\star}$ and ${\Gamma _r}$, i.e.,
\begin{align}
	{\omega ^\star} = {p^{ - 1}}{\rm{In}}\sum\limits_{r \in {{{\cal R}}_p}} {{e^{p{ \mathscr{P}}\left( {{\Gamma _r}} \right)}}}  - {p^{ - 1}}.
	\label{eq53}\end{align}
Substituting \eqref{eq53} into \eqref{eq51} to eliminate the variable ${\omega}$, yields
\begin{align}
	{\phi _p}\left( {{\Gamma _r}} \right): & = {\Psi _p}(\omega^\star,\left\{{\Gamma _r}\right\})\notag\\&
	= {p^{ - 1}}{\rm{In}}\sum\limits_{r \in {{{\cal R}}_p}} {{e^{p{ \mathscr{P}}\left( {{\Gamma _r}} \right)}}}\notag \\&
	\kern 10pt  + {p^{ - 1}}\sum\limits_{r \in {{{\cal R}}_p}} {{{\boldsymbol{1}}^T}} {e^{p\left[ {{\rm{vec}}\left( {{{\bf{\Xi }}_{\rm{r}}}} \right){\rm{ - vec}}\left( {{{\bf{\Sigma }}_{\rm{r}}}} \right){\Gamma _{\rm{r}}}} \right] - 1}}.
	\label{eq54}\end{align}
Thus, Theorem 1 follows.

\ifCLASSOPTIONcaptionsoff
\newpage
\fi

%

\bibliographystyle{IEEEtran}
\bibliography{IEEEabrv,REF}

\end{document}